\documentclass[fleqn,usenatbib]{rasti}

\usepackage{newtxtext,newtxmath}

\usepackage[T1]{fontenc}
\DeclareRobustCommand{\VAN}[3]{#2}
\let\VANthebibliography\thebibliography
\def\thebibliography{\DeclareRobustCommand{\VAN}[3]{##3}\VANthebibliography}

\usepackage{graphicx}	
\usepackage{amsmath}	
 
\usepackage{amssymb}
\usepackage{multicol}        
\usepackage{bm}		
\usepackage{pdflscape}	
\usepackage{float}
\usepackage{xurl}

\title[RFSoC receiver calibration system]{RFSoC receiver calibration system for 21-cm global spectrum experiments from space: The CosmoCube case}

\author[Jc Zhu et al.]{
Jiacong Zhu,$^{1,2}$\thanks{E-mail: zhujiacong@nao.cas.cn}
Eloy de Lera Acedo,$^{3,4}$\thanks{E-mail: ed330@cam.ac.uk}
Kaan Artuc$^{3,4}$
and Xuelei Chen$^{1,2,5}$\thanks{E-mail: xuelei@cosmology.bao.ac.cn}
\\
$^{1}$National Astronomical Observatory, Chinese Academy of Science, 20A Datun Road, Beijing 100101, China\\
$^{2}$University of Chinese Academy of Sciences, Beijing 100049, China\\
$^{3}$Cavendish Astrophysics, University of Cambridge, Cambridge, CB3 0HE, United Kingdom\\
$^{4}$Kavli Institute for Cosmology, University of Cambridge, Cambridge, CB3 0HE, United Kingdom\\
$^{5}$Department of Physics, College of Sciences, Northeastern University, Shenyang 110819, China
}

\date{Accepted 18/12/2024. Received 29/10/2024; in original form 10/05/2024}

\pubyear{\the\year{}}
\begin{document}
\label{firstpage}
\pagerange{\pageref{firstpage}--\pageref{lastpage}}
\maketitle

\begin{abstract}
The CosmoCube project plans to deploy a global 21-cm spectrometer with 10-100 MHz observation band in a lunar orbit. The farside part of such an orbit, i.e. the part of orbit behind the Moon, offers an ideal site for accurately measuring the 21-cm signal from the Dark Ages, Cosmic Dawn and Epoch of Reionization, as the effects of the Earth's ionosphere, artificial radio frequency interference (RFI), and complex terrain and soil are all avoided. Given the limitations of a satellite platform, we propose a receiver calibration system design based on a Radio Frequency system-on-chip, consisting of a Vector Network Analyzer (VNA) sub-system, and a source switching sub-system. We introduce the measurement principle of the VNA, and discuss the effect of quantization error. The accuracy, stability and trajectory noise of the VNA are tested in laboratory experiments. 
We also present the design of the source-switching sub-system, generating mock datasets, showing that the imperfect return loss, insertion loss, and isolation of surface-mounted microwave switches have a minimal effect on the sky foreground fitting residuals, which are within $\pm10$ mK under optimal fitting condition. When all possible measurement errors in reflection coefficients and physical temperatures are taken into account, the foreground fitting residuals for the 50-90 MHz part of the spectrum remain around $\pm20$ mK.
\end{abstract}




\section{Introduction}

The 21-cm spectral line produced by the hyperfine transition of neutral hydrogen is a good tracer in the study of Dark Ages, Cosmic Dawn and Epoch of Reionization. We try to measure it at low radio frequencies because of cosmological redshift. One 21-cm probe is the sky-averaged signal measured with a wide beam antenna. This is a measurement of great challenge because the Galactic synchrotron radiation foreground is about $\sim$4-5 orders of magnitude stronger. There are many ongoing ground-based experiments, such as EDGES \citep{bowman2018absorption}, SARAS3 \citep{singh2022detection}, LEDA \citep{price2018design}, PRIZM \citep{philip2019probing}, REACH \citep{de2022reach} and many other experiments. However, the observation frequency for most ground experiments cannot go below $\sim$30-40 MHz due to the ionosphere absorption. Additionally, spectrum measurement above 30 MHz will also be affected by ionosphere opacity and distortion. Moreover, artificial radio frequency interference (RFI) and complex ground and soil conditions will distort the spectrum.

Lunar orbit is an ideal site which is devoid of these issues. The moon can help shield the RFI from earth \citep{mckinley2012low}, and the strong solar emission caused by flares and coronal activity \citep{mercier1997coronal}. The absence of the ionosphere allows observation frequencies to go below $\sim$30-40 MHz, which corresponds to the Dark Ages. In space, There is no spectrum distortion caused by ionosphere, ground or soil. There are a number of proposals of lunar-based space 21-cm global spectrum experiments. \citet{Burns_2012} proposed a global spectrum measurement mission called the Dark Ages Radio Explorer (DARE), it is designed to observe 21-cm global spectrum above the lunar far side over 40 to 120 MHz with electrically-short, by-conical dipole antennas. More recently, \citet{burns2021global} proposed another mission concept called Dark Ages Polarimeter PathfindER (DAPPER), which will use a SmallSat with a total volume of $< 1\mathrm{m}^3$ to observe global spectrum over the 10-110 MHz band on orbit above the lunar far side. \citet{chen2019discovering} proposed the Discovering Sky at the Longest Wavelengths (DSL), also known as the Hongmeng project\citep{chen2023}, planning to launch an array of satellites into lunar orbit. Most of them will form a linear array making interferometric observations below 30 MHz, another one satellite will make high precision global spectrum measurement above 30 MHz. \citet{PRATUSH} proposed the Probing ReionizATion of the Universe using Signal from Hydrogen (PRATUSH), aiming to precisely measure the global 21-cm spectrum over 40 to 200 MHz in lunar orbit with a mono-cone antenna and a 2$\times$12U bus containing a radio frequency system and a digital receiver.

CosmoCube is a project led by the University of Cambridge, to perform global 21-cm spectrum measurement experiment over the 10-100 MHz using standardized CubeSat to help study the Dark Ages and Cosmic Dawn from the far side of lunar orbit \citep{artuc2024spectrometer}. CubeSats are a class of satellites that are standardized in terms of size, mechanical design, electrical design, and operational design. Such standardization significantly reduces both the financial and time costs, enabling space science research to be conducted efficiently. CosmoCube plans to use a 12U CubeSat, which is about 230-mm $\times$ 230-mm $\times$ 360-mm large, to deliver the instrument. To fit within the CubeSat platform, the instrument must be as lightweight and compact as possible, and must have a very low power consumption. 

The CosmoCube instrument consists a foldable antenna mounted on the exterior of the satellite; an radio frequency (RF) analog part including the main receiver and calibration system; and a digital receiver for sampling analog signals, performing a fast Fourier Transform (FFT) to get the spectra, and then store the data for later transmission, etc. Since the RF system and digital receiver will all be enclosed within the space of a 12U CubeSat, CosmoCube uses a Radio Frequency System-on-Chip (RFSoC) as the core component for system design. The RFSoC integrates the analog to digital converters (ADCs), the Field Programmable Gate Array (FPGA) used for digital data processing, Central Processing Unit (CPU) and several peripheral interfaces on a single board. The highly integrated solution can help reduce the size, power consumption and complexity of the system, providing high performance and flexibility for various RF applications. In addition, due to the volume constraint of the CubeSat, we need to use surface-mounted devices to form the receiver, potentially sacrificing its performance. This paper describes our design for the calibration system, which consists of a Vector Network Analyzer (VNA) sub-system and a source switching sub-system as well as their performance simulations and tests.

This paper is organized as follows. Section~\ref{sec:digital} introduces the digital performance of RFSoC. Section~\ref{sec:cal_method} describes the calibration method using noise wave parameters. Section~\ref{recevier} first describes the VNA design, tests the performance of the VNA through simulation and laboratory tests, then describes the source switching sub-system design, discusses the effect of surface-mounted devices and measurement errors on sky temperature recovery and foreground fitting residuals. Section~\ref{sec:conclusion} provides concluding remarks.

\section{Digital Receiver}
\label{sec:digital}
In this instrument, the digital board we use is a Zynq UltraScale+ RFSoC ZCU111 Evaluation Board from Xilinx. This RFSoC (XCZU28DR-2FFVG1517E) integrates eight 12-bit RF analog-to-digital converter (ADC) channels with a maximum sampling rate of 4.096 Gigabit Samples Per Second (GSPS), eight 14-bit RF digital-to-analog converter (DAC) channels with a maximum sampling rate of 6.554 GSPS, an FPGA fabric and two processors. The board offers only RFMC connectors for ADCs and DACs inherently, but the FMC-XM500 RFMC balun add-on board mates with the ZCU111 board, providing SMA connectors to facilitate user development. However, the balun board only provides two ADCs and two DACs setting for low frequency (10-3000 MHz) with 1:2 transformer, which may require future modification to meet the requirements of the CosmoCube project.

A document from the manufacturer shows that the full-scale input of the ADC is +1 Vpp, corresponding to +4 dBm, for a 50 $\Omega$ impedance system \citep{RFdataConverter}. 
Some tests characterizing the performance of the DAC and ADC have been conducted\citep{artuc2024spectrometer}. The DAC has two output modes: 32 mA and 20 mA. Here we adopt the 20 mA mode, although the maximum power will be smaller, the Spurious Free Dynamic Range (SFDR) including harmonics is about 71 dBc in single tune measurements over the frequency range of 10-100 MHz, which is better than the 63 dBc in the 32mA mode. 
When we use the DAC of the RFSoC to generate continuous wave signals at varying output power levels and receive these signals with a spectrum analyzer (a Keysight PNA-X), we observe that if the output power level exceeds -5 dBm, there is a noticeable increase in the noise floor and many spurious harmonics appear. Therefore, in designing the VNA sub-system, to ensure the highest possible signal-to-noise ratio, -5 dBm is the maximum power for measurement. 
As noted in \citet{artuc2024spectrometer}, the spectrometer design based on RFSoC can achieve an SFDR of 75dBc and an Effective Number of Bits of 11.3 bits over the frequency range of 10-100 MHz. This performance meets the requirements of low-frequency measurements where the sky temperature is approximately 30 dB higher than the ambient temperature.

\section{Calibration Methodology}
\label{sec:cal_method}
Since the 21-cm signal we aim to detect is extremely weak, approximately on the order of 100 mK, it is necessary to accurately model the system response and systematic noise. The overall design of the receiver calibration system is based on the REACH receiver introduced in \citet{razavighods2023receiver}, incorporating calibration technique developed in \citet{dicke1946measurement} along with noise wave parameters modeling of Low Noise Amplifier (LNA) \citep{meys1978wave, rogers2012absolute}. 

First-order calibration of the system can be achieved by using a Dicke switch \citep{dicke1946measurement} to make power spectral density (PSD) measurement on antenna ($P_{\mathrm{source}}$), a noise source ($P_{\mathrm{L+NS}}$) and an ambient 50 $\Omega$ load ($P_{\mathrm{L}}$) at a fixed reference plane to the receiver. A preliminary antenna temperature $T_{\text{source}}^*$ can be calculated using
\begin{equation}  
    \begin{aligned}
    T_{\text {source}}^*=T_{\mathrm{NS}}\left(\frac{P_{\text {source }}-P_{\mathrm{L}}}{P_{\mathrm{L+NS}}-P_{\mathrm{L}}}\right)+T_{\mathrm{L}}
    \end{aligned}
    \label{eq:uncal_temp}
\end{equation}
where $T_{\text{NS}}$ and $T_{\text{L}}$ represent the excess noise temperature of the noise source and the physical temperature of the ambient 50 $\Omega$ load, respectively. This method is mainly used to calibrate out time-dependent system gain arising from the components within the receiver \citep{rogers2012absolute}.
Since the noise source and the 50 $\Omega$ load are well matched (the reflection coefficients of them are typically on the order of 0.005 or less, i.e., the reflected temperature is less than 10 mK.), we can assume that the signal reflection due to the mismatch can be ignored \citep{razavighods2023receiver}.

The PSDs for them can be written as:
\begin{equation}  
    \begin{aligned}
    P_{\mathrm{L}}=g_{\text {sys }}\left[T_{\mathrm{L}}\left(1-\left|\Gamma_{\text {rec }}\right|^2\right)+T_0\right]
    \end{aligned}
    \label{eq:P_L}
\end{equation}
and
\begin{equation}  
    \begin{aligned}   P_{\mathrm{L+NS}}=g_{\mathrm{sys}}\left[\left(T_{\mathrm{L}}+T_{\mathrm{NS}}\right)\left(1-\left|\Gamma_{\mathrm{rec}}\right|^2\right)+T_0\right]
    \end{aligned}
    \label{eq:P_NS}
\end{equation}

Here $\Gamma_{\mathrm{rec}}$ is the reflection coefficient of the receiver, $g_{\text{sys}}$ and $T_{\text{0}}$ are the system gain and receiver noise offset respectively.

When considering other sources with non-negligible reflection coefficients, the PSD function can be written as \citep{monsalve2017calibration}
\begin{equation}  
    \begin{aligned}
    P_{\text{source}}=&g_{\text{sys}}\left[T_{\text {source}}\left(1-\left|\Gamma_{\text {source}}\right|^2\right)\left|\frac{\sqrt{1-|\Gamma_{\text{rec}}|^2}}{1-\Gamma_{\text{source}}\Gamma_{\text{rec}}}\right|^2\right. \\
    & +T_{\text{unc}}\left|\Gamma_{\text {source}}\right|^2\left|\frac{\sqrt{1-|\Gamma_{\text{rec}}|^2}}{1-\Gamma_{\text{source}}\Gamma_{\text{rec}}}\right|^2 \\
    & +T_{\cos} \text{Re}\left(\Gamma_{\text {source}} \frac{\sqrt{1-|\Gamma_{\text{rec}}|^2}}{1-\Gamma_{\text{source}}\Gamma_{\text{rec}}}\right) \\
    & \left.+T_{\sin} \text{Im}\left(\Gamma_{\text {source}} \frac{\sqrt{1-|\Gamma_{\text{rec}}|^2}}{1-\Gamma_{\text{source}}\Gamma_{\text{rec}}}\right)+T_{\text{0}}\right] . \\
    &
    \end{aligned}
    \label{eq:nw_psd}
\end{equation}
Here, 
$T_{\text{source}} $ is the calibrated input temperature, $\Gamma_{\text{source}}$ is the reflection coefficient of the antenna, and $\Gamma_{\text{rec}}$ is the reflection coefficient of the receiver. $T_{\text{unc}}$, $T_{\text{cos}}$, $T_{\text{sin}}$ are the noise wave parameters of LNA introduced by \citet{meys1978wave}. $T_{\text{unc}}$ represents the portion of input noise that is uncorrelated with the noise at the LNA output, while $T_{\text{cos}}$ and $T_{\text{sin}}$ are components of the correlated portion. 

Substituting the expressions for $P_{\mathrm{L}}$, $P_{\mathrm{L+NS}}$ and $P_{\mathrm{source}}$ into the equation~(\ref{eq:uncal_temp}), we can get the calibration equation
\begin{equation}  
    \begin{aligned}
    T_{\mathrm{NS}}\left(\frac{P_{\text {source }}-P_{\mathrm{L}}}{P_{\mathrm{L+NS}}-P_{\mathrm{L}}}\right)+T_{\mathrm{L}} & =T_{\text {source }}\left[\frac{1-\left|\Gamma_{\text {source }}\right|^2}{\left|1-\Gamma_{\text {source }} \Gamma_{\text {rec }}\right|^2}\right] \\
    & +T_{\text {unc }}\left[\frac{\left|\Gamma_{\text {source }}\right|^2}{\left|1-\Gamma_{\text {source }} \Gamma_{\text {rec }}\right|^2}\right] \\
    & +T_{\cos }\left[\frac{\operatorname{Re}\left(\frac{\Gamma_{\text {source }}}{1-\Gamma_{\text {source }} \Gamma_{\text {rec }}}\right)}{\sqrt{1-\left|\Gamma_{\text {rec }}\right|^2}}\right] \\
    & +T_{\sin }\left[\frac{\operatorname{Im}\left(\frac{\Gamma_{\text {source }}}{1-\Gamma_{\text {source }} \Gamma_{\text {rec }}}\right)}{\sqrt{1-\left|\Gamma_{\text {rec }}\right|^2}}\right]
    \end{aligned}
    \label{eq:cal_equation}
\end{equation}
which can be simplified as \citep{roque2021bayesian}
\begin{equation}  
    \begin{aligned}
    T_{\text {source }}=X_{\text {unc }} T_{\mathrm{unc}}+X_{\cos } T_{\cos }+X_{\sin } T_{\mathrm{sin}}+X_{\mathrm{NS}} T_{\mathrm{NS}}+X_{\mathrm{L}} T_{\mathrm{L}}
    \end{aligned}
    \label{eq:nw_psd_simple}
\end{equation}

Using multiple calibrators with different impedance and measuring their reflection coefficients and PSDs, we can obtain the measured quantities (X-terms). With the measured physical temperatures of these sources, we can calculate noise wave parameters $T_{\text{unc}}$, $T_{\text{cos}}$, $T_{\text{sin}}$, $T_{\text{NS}}$ and $T_{\text{L}}$, then recover the sky temperature.

\section{Receiver Calibration system}
\label{recevier}
Observations will be conducted on the far side of the lunar orbit, where the change of spacecraft temperature may affect the antenna reflection coefficient. Although the satellite will be equipped with an internal temperature controller, the receiver will inevitably experience temperature drift. So it is necessary to measure the reflection coefficients of different components and calculate noise wave parameters of the LNA in real-time for precise system calibration. 

\begin{figure*}
	\includegraphics[width=2\columnwidth]{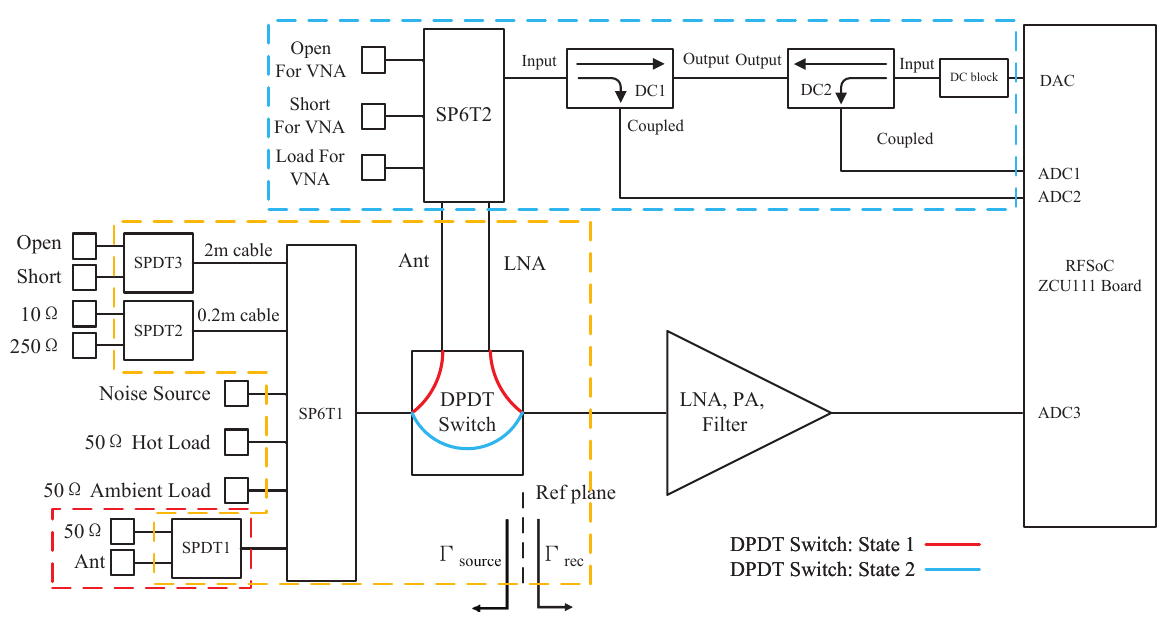}
    \caption{The block diagram of the analog receiver. The part framed by dashed blue line is the VNA sub-system. 'DC1' and 'DC2' are two directional couplers that constitute the VNA-subsystem. 'SP6T' is the 6-way microwave switch. 'SPDT' is the 2-way microwave switch. 'DPDT' is the double pole double throw switch. $\Gamma_{\text{source}}$ represents the reflection coefficient of the antenna or calibrators, $\Gamma_{\text{rec}}$ is the reflection coefficient of the receiver. A microwave switch is used in the dashed red box to enhance isolation and reduce antenna signal leakage. The part framed by dashed orange line is considered as a multi-port microwave network in the mock datasets generation process. The solid red and blue lines represent the signal paths for the different states of the DPDT switch. When the DPDT switch is in state 1, the VNA measures $S_{\text{11}}$ of the antenna, calibrators and the LNA. When the DPDT switch is in state 2, the RFSoC measures the spectrum of antenna and calibrators }
    \label{fig:receiver_block_diagram}
\end{figure*}

Figure~\ref{fig:receiver_block_diagram} shows the block diagram of the calibration system in the analog receiver we will use for simulation and test. The calibration system mainly consists of a VNA sub-system detailed in section~\ref{sec:VNA}, and a source switching sub-system introduced in section~\ref{sec:source_switch}. We performed simulations to evaluate the measurement accuracy of the receiver. The simulation method is outlined in section~\ref{sec:simulation}, with the results discussed in section~\ref{sec:results}.

\subsection{Vector Network Analyzer Sub-system}
\label{sec:VNA}
\subsubsection{ Measurement principles}

\begin{figure}
	\includegraphics[width=\columnwidth]{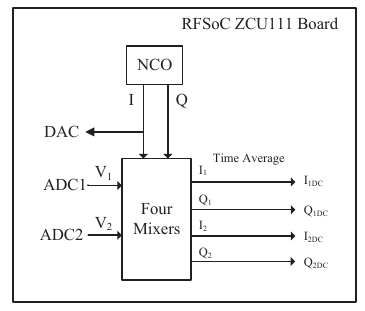}
    \caption{The internal computation process of the RFSoC FPGA.}
    \label{fig:VNA_calculation}
\end{figure}

We used the DAC and ADC of the RFSoC to build the VNA measurement circuit, as shown within the dashed blue line in Figure~\ref{fig:receiver_block_diagram}. Figure~\ref{fig:VNA_calculation} shows the internal algorithms of the RFSoC FPGA. A Numerically Controlled Oscillator (NCO) in the FPGA is used to generate a single-frequency cosine signal $I = \cos{(2\upi ft)}$ and a sine signal $Q = \sin{(2\upi ft)}$,  where $f$ is the frequency and $t$ is the time. The signal frequency sweeps across the entire observation band, which is 10-100 MHz for our instrument. The cosine signal is converted to analog via a DAC, filtered with a direct current (DC) block and passed two cascaded directional couplers before being reflected by the device under test (DUT). The signal flowing out from the coupled port of the first directional coupler DC1 is a portion of the cosine signal generated by the NCO, denoted as $V_1 = A_1\cos{(2\upi ft+\phi_1)}$. The signal flowing out from the coupled port of the second directional coupler DC2 is a portion of the signal reflected by the DUT, denoted as $V_2 = A_2\cos{(2\upi ft+\phi_2)}$. They will be sampled by two ADCs of RFSoC and then mixed with the I, Q signal, resulting in four signals $I_1, Q_1, I_2, Q_2$:
\begin{equation}
    \begin{aligned}
    I_1 &= IV_1 = \frac{1}{2}A_1[\cos(4\upi ft+\phi_1)+\cos(\phi_1)]\\
    Q_1 &= QV_1 = \frac{1}{2}A_1[\sin(4\upi ft+\phi_1)-\sin(\phi_1)]\\
    I_2 &= IV_2 = \frac{1}{2}A_2[\cos(4\upi ft+\phi_2)+\cos(\phi_2)]\\
    Q_2 &= QV_2 = \frac{1}{2}A_2[\sin(4\upi ft+\phi_2)-\sin(\phi_2)]
    \end{aligned}
	\label{eq:VNA_mixer}
\end{equation}

Next, we extract the DC components of these four signals by time-averaging, resulting in four DC signals,
\begin{equation}
    \begin{aligned}
    I_{1DC} &= \frac{1}{2}A_1\cos(\phi_1)\\
    Q_{1DC} &= -\frac{1}{2}A_1\sin(\phi_1)\\
    I_{2DC} &= \frac{1}{2}A_2\cos(\phi_2)\\
    Q_{2DC} &= -\frac{1}{2}A_2\sin(\phi_2)
    \end{aligned}
	\label{eq:VNA_average}
\end{equation}

If $V_1,V_2$ represent the incident and reflected voltages of the DUT, the reflection coefficient of DUT can be expressed as :

\begin{equation}
    \Gamma = \frac{A_2}{A_1}e^{i(\phi_2-\phi_1)} = \frac{I_{2DC}-iQ_{2Dc}}{I_{1DC}-iQ_{1Dc}}
	\label{eq:VNA_raw}
\end{equation}

In the practical case with non-ideal directional couplers, $V_1$ and $V_2$ are not exactly the incident and reflected voltages of the DUT, and the raw reflection coefficient $\Gamma$ is not exactly that of the DUT but needs correction. According to \citet{Doug1996}, a single-port vector measurement network can be modeled by a 3-term error model, including the directivity error $E_D$, match error $E_S$ and tracking error $E_R$. By measuring three known standards, these errors can be calculated to determine the true reflection coefficient of the DUT.

Although in principle the errors are eliminated by the VNA calibration process with the Open-Short-Load (OSL) standards, the quantization errors arising from analog-to-digital conversion could lead to a loss of information. This is particularly noticeable when measuring a 50 $\Omega$ load with a very low reflection coefficient or when measuring active components such as LNA with low power. For surface-mounted directional couplers, which is used because of the limited space of the CubeSat, but have higher losses, larger reflection coefficients, and poorer directivity than coaxial directional couplers, this error can be even more significant. In the following subsections, we will investigate the above issues in the RFSoC-based VNA design through simulations and experimentation to assess the measurement accuracy.

\subsubsection{Quantization error simulation}

We conducted simulation tests using the MATLAB Simulink Platform. A program of VNA measurement simulation built based on the SoC Blockset and SoC Blockset Support Packages for the Xilinx Devices in the MATLAB Simulink library. If the simulation results meet expectations, we can directly generate Hardware Description Language code and deploy it to the RFSoC for laboratory tests, which will be introduced in Section~\ref{sec:vna_test}

The configuration of RFSoC Data Converter should be the same for both simulation and laboratory tests. The stream clock frequency is 256 MHz. 4 NCOs are utilized to generate 16-bit signals with the same frequency and adjacent phases, ensuring that the DAC outputs four samples per clock cycle. Similarly, the ADC get 4 samples per clock cycle. The interpolation factor for DAC and the decimation factor for ADC are both 4. Based on the configuration above, the sampling rate of ADCs and DACs are 4.096 GSPS. 
The NCO takes a phase increment as input and accumulates it, using the quantized output of accumulator as an index to the Lookup Table of the sine wave values and then produce accurate single frequency signal. To implement frequency sweeping functionality, a phase increment value is needed by the NCO to generate signals at varying frequencies. This is given by 
\begin{equation}
    \Delta \Phi = \text{round}(2^Nf/S)
	\label{phase_increament}
\end{equation}
where $f$ is the output frequency, $N$ is the length of accumulator word length, $S$ is the sampling rate. Due to the phase increment being an integer, the frequency resolution is constrained by the sampling rate and accumulator word length. In this case, the sampling rate, $S$ is $4 \times 256 \text{ MHz} = 1024 \text{ MHz}$. To achieve a frequency resolution better than 0.1 MHz, the word length is set to 14. Additionally, we configure the dither bit of NCOs to add some dithering noise to help convert quantized noise to white noise, reducing harmonics and spurious signals. The cost of this is a lower signal-to-noise ratio. We tested different values, and chose a trade-off value of 4.

After the ADC samples the RF signal and converts it to a digital signal, the FPGA internally executes the mixing process described in equation~(\ref{eq:VNA_mixer}) and the averaging process described in equation~(\ref{eq:VNA_average}). Subsequently, the data is transferred via the AXI bus into the Programmable Logic module for further computation and storage. In both simulation and laboratory tests, these data will be imported into MATLAB workspace for raw reflection coefficient calculation and calibration.

In simulation, we assume that the reflection coefficients (i.e. $S_{\text{11}}$) of the Open, Short, Load calibration standards are constant across the entire frequency band, with respective values of 1, -1, and 0. We choose two different sources, representing the high and low reflection coefficients. Source 1 has an $S_{\text{11}}$ magnitude of 0.5, while that of source 2 is 0.02, which is close to the LNA used in the receiver. The phase of both varies linearly with frequency. Additionally, We used several power levels to study the effects of quantization error, including -5 dBm, -20 dBm, -25 dBm, and -30 dBm. We implemented two different calibration methodologies: one is to calibrate with -5 dBm using OSL calibration standards, and then measure the DUTs with varing power levels; the other is calibrating and measuring both with the same power level. To speed up the simulation, the frequency sweep interval is set to be 1 MHz.

\begin{figure}
	\includegraphics[width=\columnwidth]{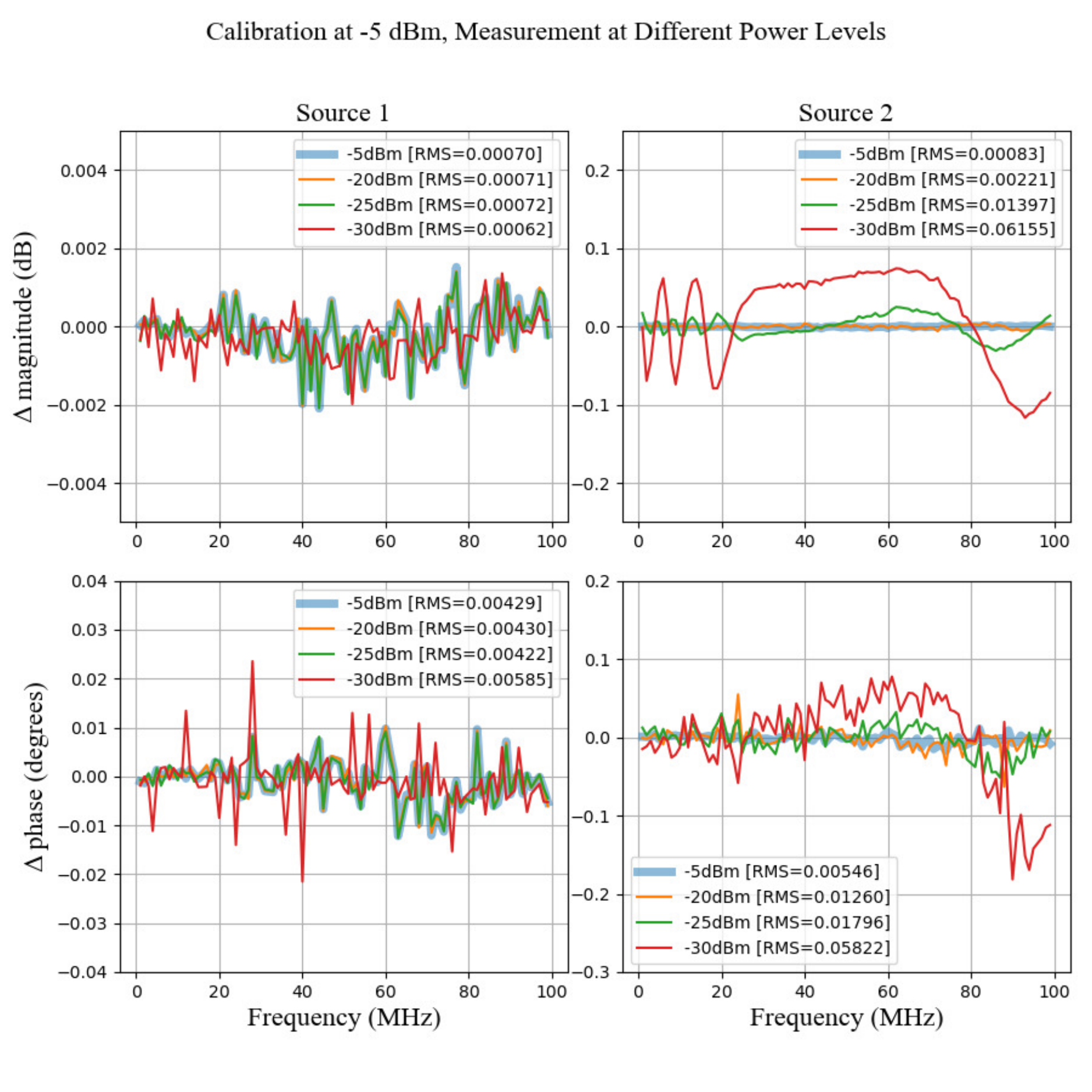}
    \caption{Measurement errors in VNA simulation when calibrating with a -5 dBm power level and measuring with different power levels: The first column corresponds to source 1 with $\Gamma_1 = 0.5\exp{(-i\times0.01f)}$, and the second column to source 2 with $\Gamma_2 = 0.02\exp{(-i\times0.01f)}$.  The first row represents the magnitude errors in dB, and the second row represents the phase errors in degrees. Different colored lines represent different power levels. The root mean squares (RMS) of the measurement errors are labelled in the legend.}
    \label{fig:cal_meas_diff}
\end{figure}

\begin{figure}
	\includegraphics[width=\columnwidth]{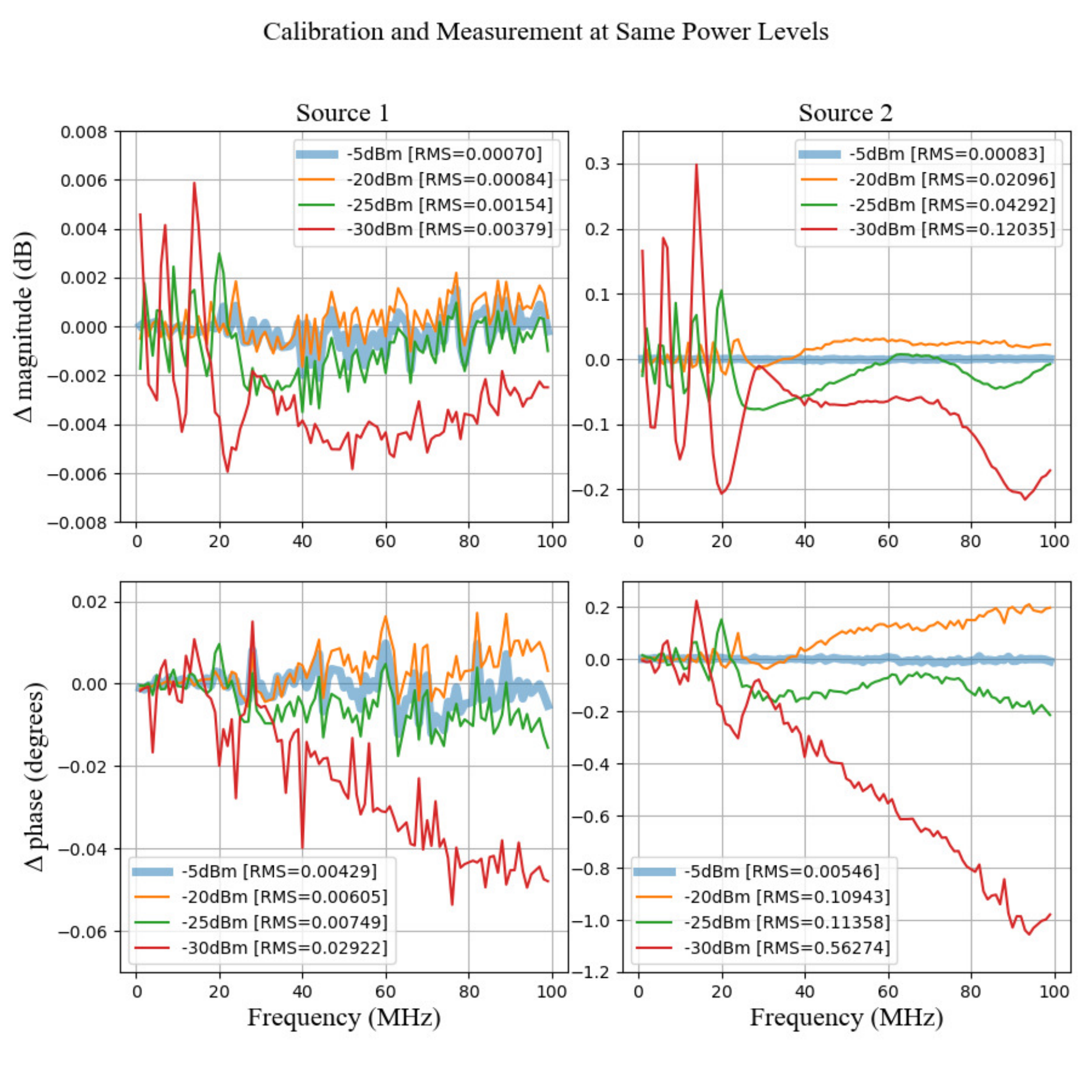}
    \caption{Measurement errors in VNA simulation when calibrating and measuring with the same power level. The first column corresponds to source 1 and the second column to source 2. Different colored lines represent different power levels. The RMS of the measurement errors are labelled in the legend.}
    \label{fig:cal_meas_same}
\end{figure}

\begin{figure}
	\includegraphics[width=\columnwidth]{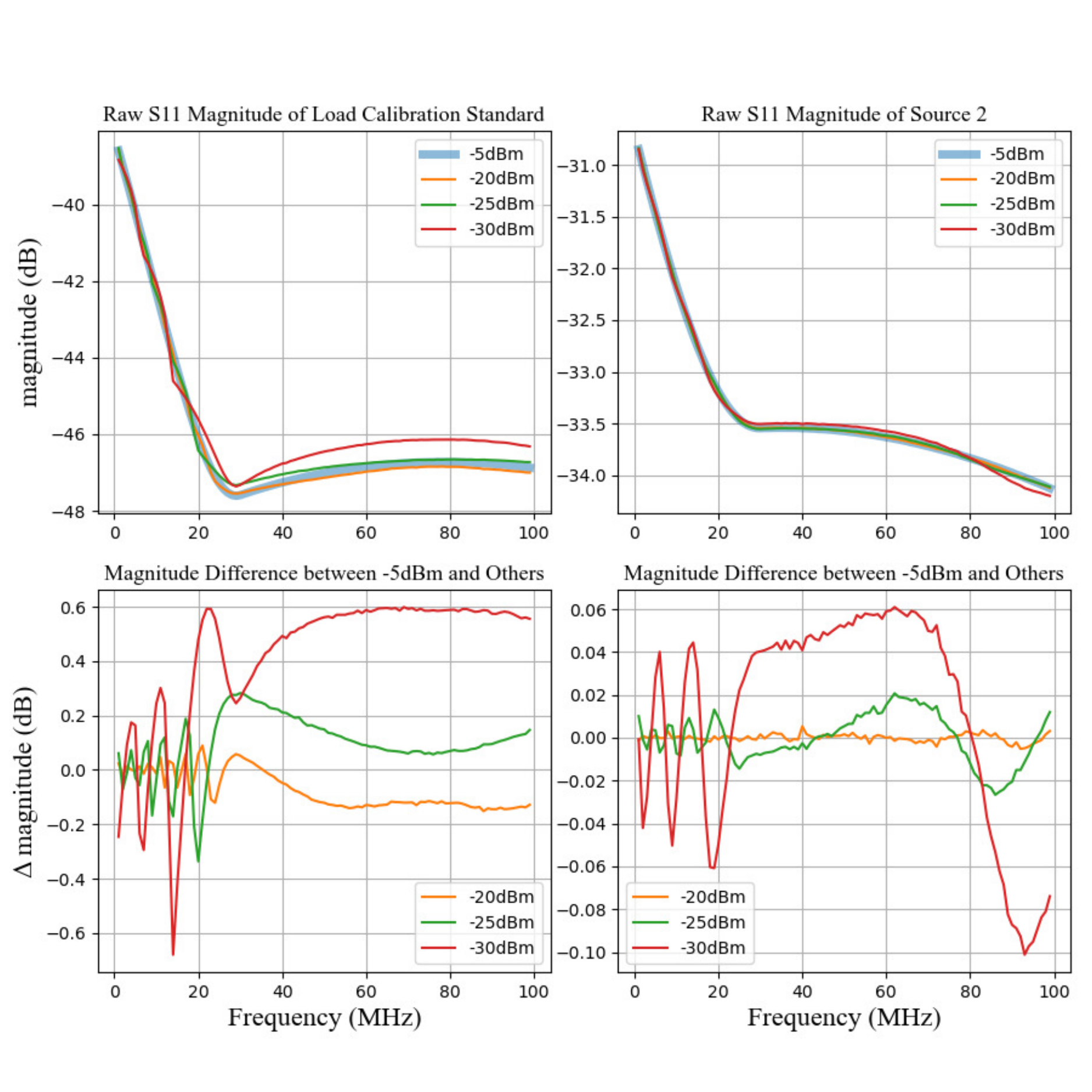}
    \caption{Top panel shows the magnitude of raw measurement $S_{\text{11}}$. Bottom panel shows the magnitude difference between the raw $S_{\text{11}}$ obtained from the low-power measurements and the -5 dBm measurement. Left: Load Calibration Standards. Right: Source 2.}
    \label{fig:raw_s11_diff}
\end{figure}

Figure~\ref{fig:cal_meas_diff} shows the measurement errors when calibrating with -5 dBm power level and measuring with different power levels. The VNA works well when measuring a DUT with a high reflection coefficient, but it exhibits significant frequency structural errors when measuring a DUT with low reflection coefficient using low power. Figure~\ref{fig:raw_s11_diff} shows that the measurement error is mainly due to the weak reflected signal of the low reflection coefficient DUT, as the structures are fundamentally similar. Figure~\ref{fig:cal_meas_same} shows the measurement errors when calibrating and measuring with the same power levels. Errors during low-power measurement are significantly increased due to the introduction of measurement errors in the raw $S_{\text{11}}$ of the Load calibration standard.

In simulation, changes in the output power do not affect the single-port VNA error parameters. However, in practice, different output powers might alter the single-port error parameters, and the parameters of directional couplers are not completely the same as those in the simulation. The most crucial insight from the simulation is the need to be vigilant about the measurement accuracy for low reflection coefficients with low power.

\subsubsection{Performance evaluation}
\label{sec:vna_test}
We conducted laboratory tests on the RFSoC-based VNA and compared it with a commercial VNA to evaluate its practical performance. For this comparison, we utilized a N5247A PNA-X Microwave Network Analyzer, which is calibrated with the N4433A Electronic Calibration Module. In the following text, it will be referred to as 'PNA'. For the RFSoC, only a mechanical calibration kit could be used, specifically the Anritsu 3650 calibration kit in this test. To ensure the accuracy of the comparison and eliminate any potential errors from calibration kits, we adopted the following experimental approach:

\begin{enumerate}
 \item Calibrate the PNA using the N4433A Electronic Calibration Module.
 \item Measure the S-parameters of mechanical OSL calibration standards and a series of different DUTs with the PNA, treating the results as their true S-parameters.
 \item Measure the raw S-parameters of the same OSL calibration standards using the RFSoC-based VNA, and calculate the single-port error parameters.
 \item Measure the raw S-parameters of same DUTs using the RFSoC-based VNA, calibrate them with previous error parameters to obtain the calibrated S-parameters.
 \item Compare the results of the PNA and the RFSoC-based VNA.
\end{enumerate}

\begin{figure}
	\includegraphics[width=\columnwidth]{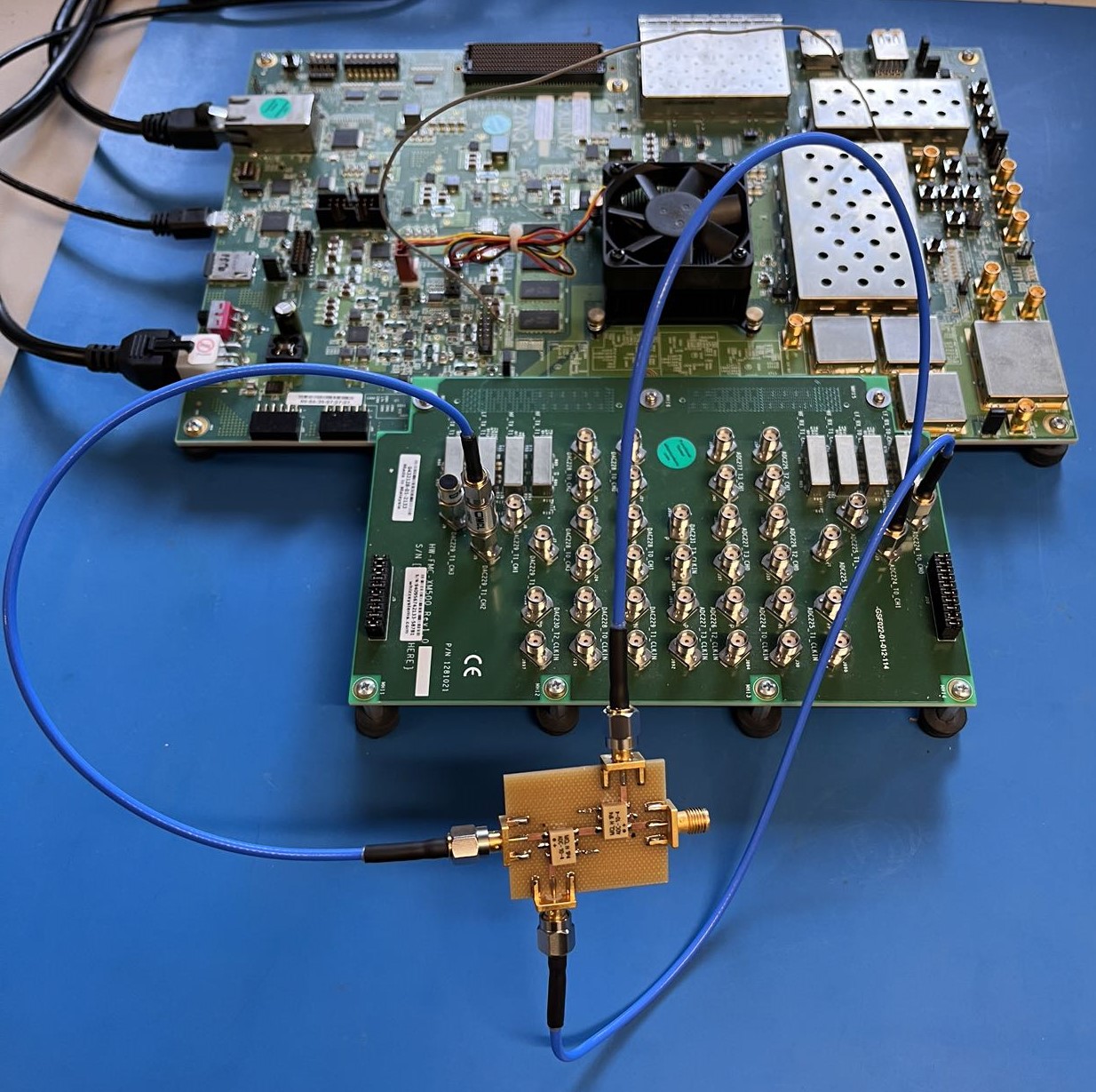}
    \caption{A photo of the test setup. The ZCU111 evaluation board is connected to the balun add-on board called XM500 providing SMA connectors. At the forefront is a circuit board comprising two ADC-10-4+ directional couplers. The input port is connected to one DAC, and the coupled ports of the two directional couplers are connected to the two ADC ports. The output port is open here and will be connected to different DUTs in the test.}
    \label{fig:Test_Setup}
\end{figure}

Figure~\ref{fig:Test_Setup} shows the test setup. We integrated two ADC-10-4+ directional couplers on a single circuit board and connected the input port and two coupling ports of this board to the DAC and ADC SMA ports on the balun add-on board. A DC blocker was connected to the DAC output on the balun add-on board to prevent the influence of DC signal. A specially designed connection scheme will be used for the satellite, while here it is just for testing purposes. 

A variety of test objects were selected as DUTs. Our choices included attenuators with variant resistances, primarily to evaluate the accuracy of the VNA when measuring $S_{\text{11}}$ with flat amplitude and phase. We also chose a narrowband filter with resonance points in their $S_{\text{11}}$ within the band to evaluate the accuracy of VNA in measuring rapidly changing amplitude and phase. Additionally, we selected a cable, as it will be used in noise parameters calculation. Following the aforementioned criteria, we selected an 1 dB attenuator, a 13 dB attenuator, a 22 MHz filter and a 10-meter cable. During measurement, one end of these DUTs is connected to an Open calibration standard, and the other end is connected to the output port of the circuit board. We used a 13 dB attenuator because its $S_{\text{11}}$ is approximately -26 dB when terminated with an Open calibration standard, which is close to the $S_{\text{11}}$ of the LNA. 

\begin{figure}
	\includegraphics[width=\columnwidth]{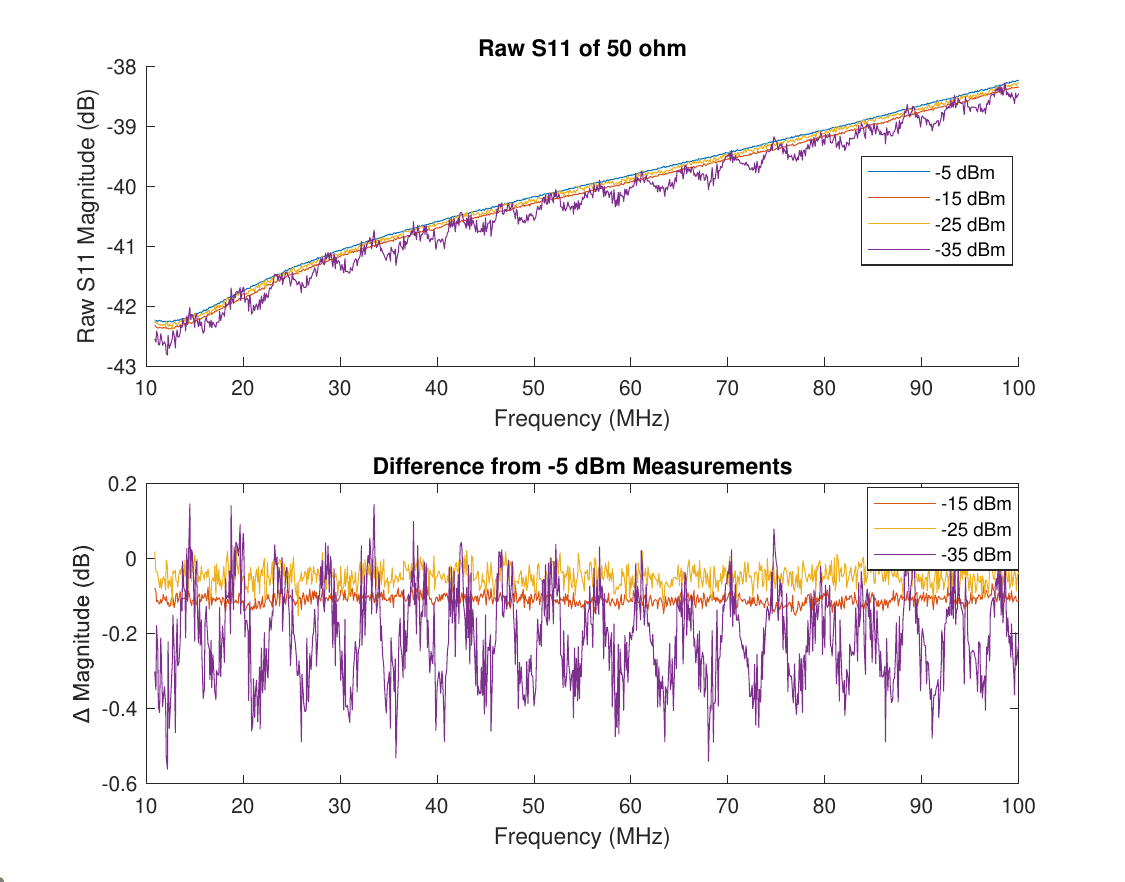}
    \caption{The raw $S_{\text{11}}$ measurement results of a 50 $\Omega$ Load Calibration standard. Top figure shows the magnitude of the  raw $S_{\text{11}}$ with different power levels. Bottom figure shows the magnitude difference between the measurement results with -5 dBm power level and with other power levels.}
    \label{fig:raw_s11_of_load}
\end{figure}

First, we examined the raw $S_{\text{11}}$ measurement results of the Load calibration standard to see if there are frequency structural errors caused by quantization errors. Figure~\ref{fig:raw_s11_of_load} compares results using different power levels, with -5 dBm as the standard. A noticeable wiggling pattern appears at measurement with -35 dBm power, and this pattern will become more obvious when using lower power, which is not shown in this figure but has been verified through experiments. Although the measurement result with -25 dBm exhibits higher noise levels,  it avoids significant structural errors, and this input power level will not damage the LNA chips. Therefore, in our subsequent tests, we have chosen -25 dBm as the low-power output.

\begin{figure*}
	\includegraphics[width=2\columnwidth]{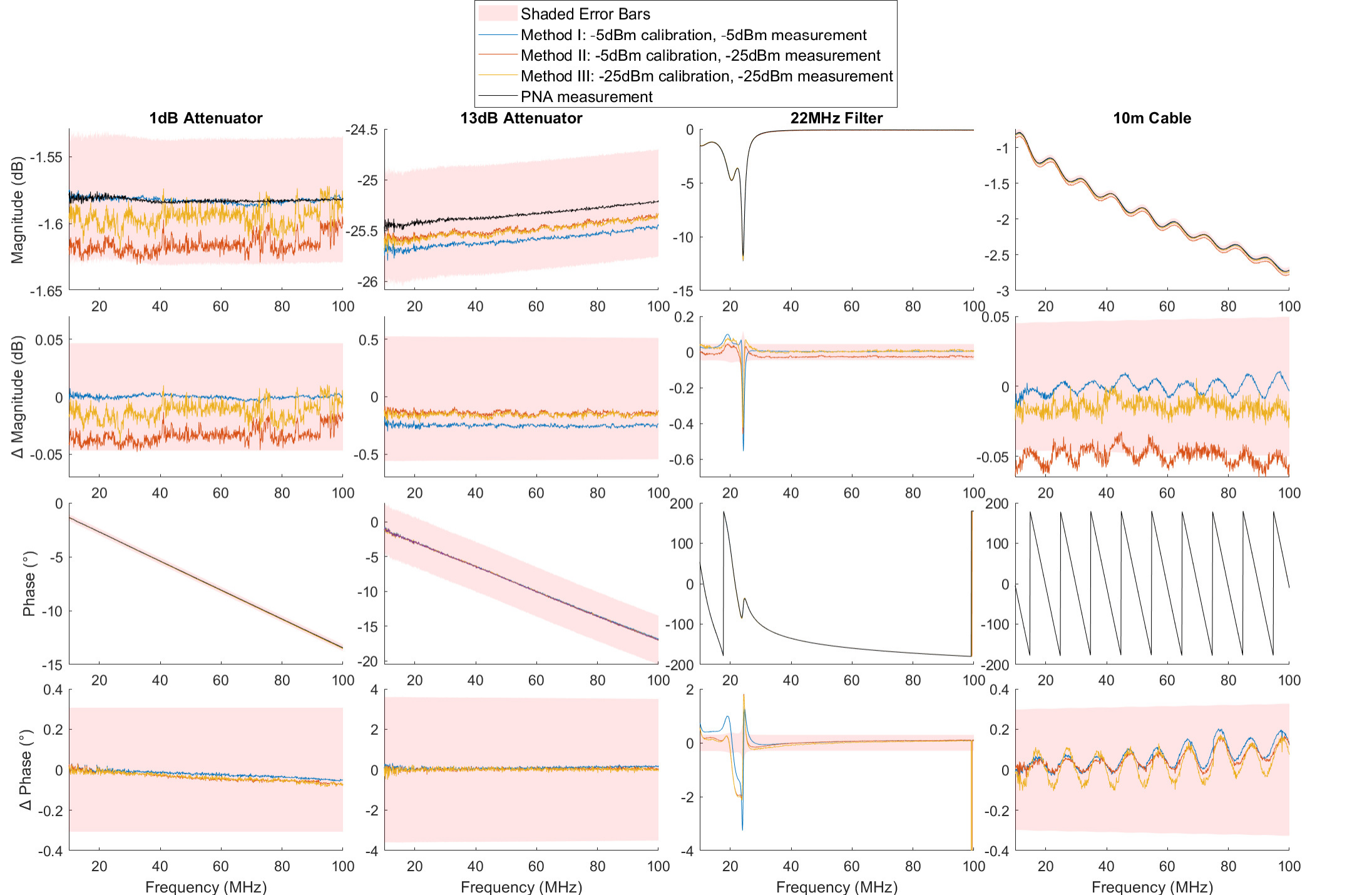}
    \caption{The measurement results of the RFSoC-based VNA and the PNA for four DUTs and their differences. The black line represents PNA measurement results, while the other different colored lines represent measurement results of different power methods for RFSoC-based VNA. The shaded area represents the measurement uncertainty of PNA. The four columns correspond to a 1 dB attenuator, a 13 dB attenuator, a 22 MHz filter, and a 10-meter cable, all of which are terminated with an Open calibration standard. The first and third rows are magnitude in dB and phase in degrees respectively. The second and fourth rows are the magnitude and phase differences from the PNA measurements, respectively.}
    \label{fig:cal_s11_diff}
\end{figure*}

In the accuracy testing of the RFSoC-based VNA, we adopted three different measurement methods, which are listed in Table~\ref{tab:vna_method}.
\begin{table}
	\centering
	\caption{Calibration power level and measurement power level used in three different measurement methods.}
	\label{tab:vna_method}
	\begin{tabular}{ccccc} 
		\hline
		Method &  Calibration Power & Measurement Power \\
		\hline
            \uppercase\expandafter{\romannumeral1} & -5 dBm & -5 dBm\\
		\uppercase\expandafter{\romannumeral2} & -5 dBm & -25 dBm\\
            \uppercase\expandafter{\romannumeral3} & -25 dBm & -25 dBm\\
            \hline
	\end{tabular}
\end{table}
In the test, the frequency resolution is 0.0675 MHz and the measurement duration is 0.01 second per frequency point. Figure~\ref{fig:cal_s11_diff} shows the measurement results of the RFSoC-based VNA and the PNA, as well as the differences between them. The shaded area represents the measurement uncertainty of PNA, obtained by the VNA uncertainty calculator provided by \citet{Keysight}. For sources with low reflection coefficients, the measurement error of RFSoC-based VNA and uncertainty of PNA are both larger, indicating more difficult measurements for sources with low reflection coefficients. 

\begin{figure}
	\includegraphics[width=\columnwidth]{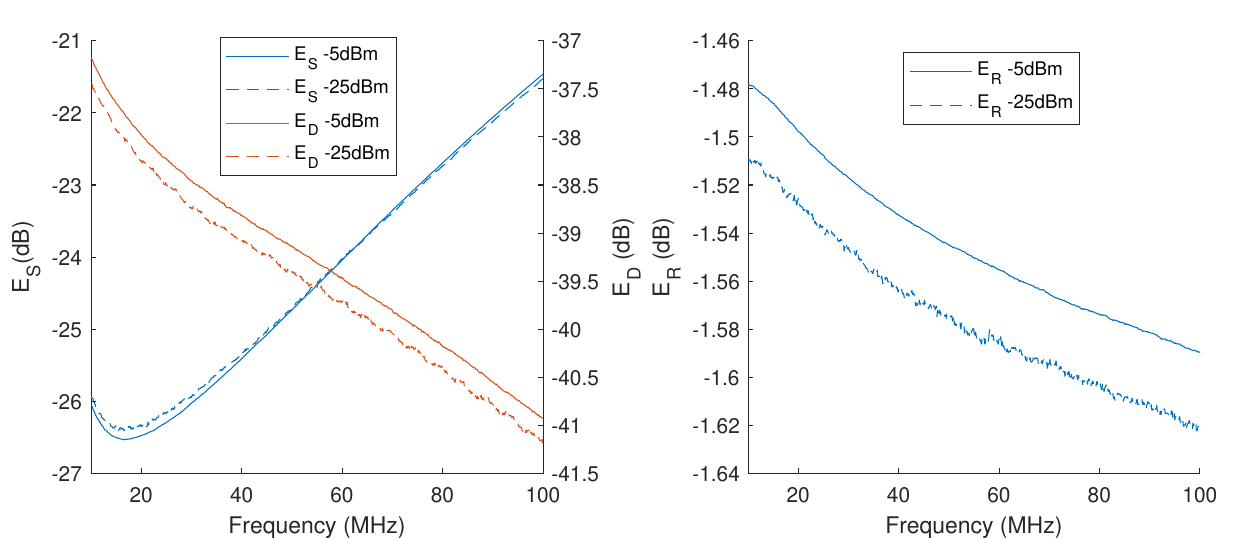}
    \caption{The error parameters of 3-term error model calculated at two power levels, -5 dBm and -25 dBm.}
    \label{fig:error_param}
\end{figure}

These results show that in the actual experiments, the accuracy of method \uppercase\expandafter{\romannumeral3} is higher than that of method \uppercase\expandafter{\romannumeral2} in most cases. For the 13 dB attenuator, although method \uppercase\expandafter{\romannumeral1} deviates the most from the PNA measurements, this is likely due to the stability of cables and connectors for low reflection coefficient measurement \citep{buber2019characterizing}. Regarding method \uppercase\expandafter{\romannumeral1} as a standard, method \uppercase\expandafter{\romannumeral3} still outperforms method \uppercase\expandafter{\romannumeral2}. It is different from the conclusions drawn in the simulation, indicating that the 3-term error parameters of single-port VNA have changed after altering the power level. Figure~\ref{fig:error_param} shows the 3-term error parameters calculated with two power levels, and slight differences can be observed. Notably, all measurement errors are within the uncertainty of the PNA, except the resonance point of the 22 MHz filter. It may be due to the rapid change of $S_{\text{11}}$ near the resonance point in logarithmic coordinates and is more likely to cause larger measurement errors, which needs to be improved in the future. In section~\ref{sec:results}, we will explore the impact of VNA measurement errors on the 21-cm sky signal by simulating the entire analogue receiver.

\begin{figure}
	\includegraphics[width=\columnwidth]{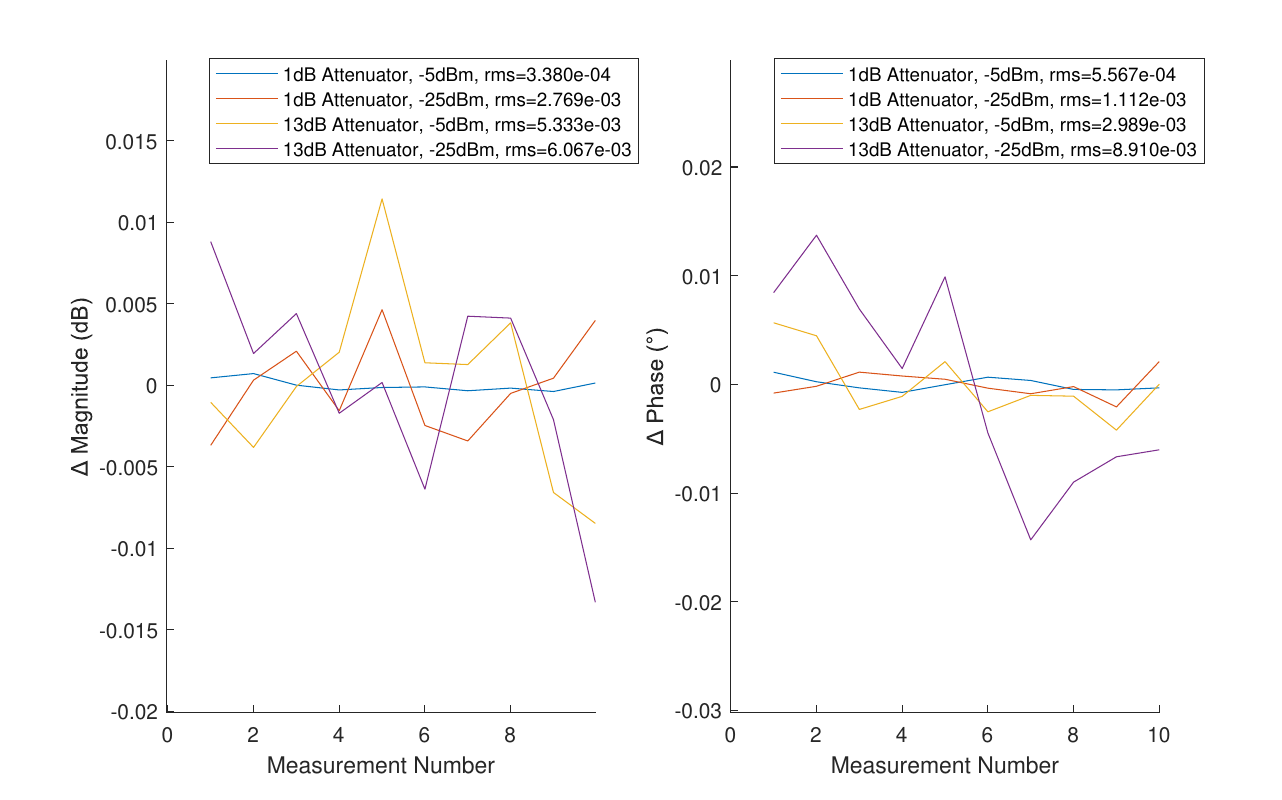}
    \caption{Continuously measure the 1 dB and 13 dB attenuators 10 times with powers of -25 dBm and -5 dBm and average the values across frequencies. Then, subtracted the average of these 10 values from their respective magnitudes and phases. The first column is Magnitude in dB, the second column is phase in degrees.}
    \label{fig:stability}
\end{figure}

\begin{figure}
	\includegraphics[width=\columnwidth]{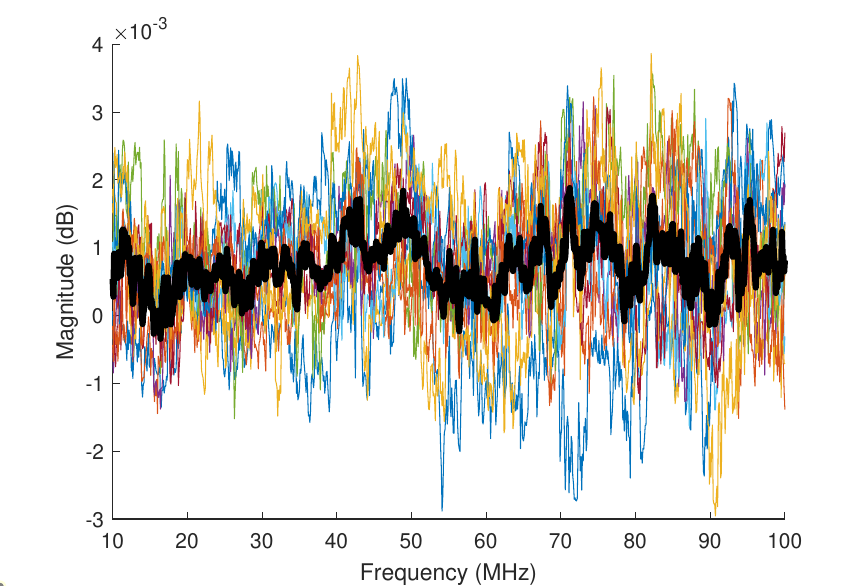}
    \caption{The results of ten continuous measurements of the Open calibration standard with -5 dBm power level; the thick black line represents the average of these ten measurements.}
    \label{fig:noise}
\end{figure}

We also conducted tests on the stability and trajectory noise. Figure~\ref{fig:stability} illustrates the stability of the RFSoC-based VNA. As expected, when measuring low reflection coefficient devices with low power, the results are the least stable, with the magnitude's RMS being less than $10^{-2}$ dB and the phase's RMS being less than $5\times10^{-3}$ degrees. Figure~\ref{fig:noise} illustrates the trace noise of VNA measurement. Here, for better visibility of the noise level, it is assumed that the Open calibration standard is ideal, with a reflection coefficient of 0 dB. The noise of a single measurement is about $\pm2$\textasciitilde$3\times10^{-3}$ dB across the frequency band, and after 10 times of averaging, the noise can be effectively reduced to less than $\pm1\times10^{-3}$ dB, much less than the measurement error showed in Figure~\ref{fig:cal_s11_diff}.

\subsection{Source Switching Sub-system}
\label{sec:source_switch}
In order to calculate the noise wave parameters of the LNA in real time, several calibrators are required within the instrument. In \citet{razavighods2023receiver}, REACH used 12 different calibrators, including a 2-m and a 10-m cable. This approach was aimed at providing frequency-dependent impedance and filling the Smith chart as much as possible. However, due to the satellite's size constraints, it is not feasible to use so many calibrators on CosmoCube. As shown in Figure~\ref{fig:receiver_block_diagram}, in addition to the noise source, We use 6 calibrators as follows:
\begin{enumerate}
 \item A hot 50 $\Omega$ load heated to 370 K.
 \item An ambient temperature 50 $\Omega$ load 
\item A 0.2-m cable connected to switch SPDT2 (terminated in 10 $\Omega$, or 250 $\Omega$) at ambient temperature.
 \item A 2-m cable connected to switch SPDT3 (terminated inn Open, or Short) at ambient temperature.
\end{enumerate}

The part framed by the dashed orange lines in Figure~\ref{fig:receiver_block_diagram} is the source switching sub-system. 
Table~\ref{tab:switch_comparison_table} compares the electrical specifications of the surface-mounted microwave switches and the mechanical switches. Insertion loss describes the signal power loss during transmission. Return loss describes the reflection of signal power due to impedance mismatch bewteen different transmission lines or devices. Isolation is the degree to which a device or component prevents unwanted signal coupling between its ports. For surface-mounted switches, the higher return loss complicates system response, necessitating more precise $S_{\text{11}}$ measurements for both the antenna and receiver. Higher insertion loss can cause ambient temperature to leak into the spectrum, requiring accurate lab measurements and real-time temperature monitoring. Poorer isolation may lead to signal leakage between paths, making it difficult to eliminate during calibration due to the challenges in accurately measuring isolation. Nevertheless, due to the limited space, the surface-mounted microwave switches will be used due to the limited space available. 

If isolation needs to be taken into account in calibration, the entire calibration process will involve a very large number of parameters, which is too complex and less robust. So we hope to ensure that signals from different paths remain independent and unaffected by each other in our design. We adopted a switch SPDT1 to reduce signal leakage from antenna. It switches to the antenna when the receiver is measuring it, and to a 50 $\Omega$ load when the receiver is measuring other calibrators. Similarly, the noise source will be switched off when it is not being measured.

\begin{table}
	\centering
	\caption{The RF electrical specifications of mechanical 6-way microwave switch MSP6TA-12+ and surface-mounted 6-way microwave switch JSW6-33DR+. All data are based on typical results from measurements taken within 5-1000 MHz, at 25$^{\circ}$C ambient temperature and under normal supply voltage, sourced from the Xilinx website.}
	\label{tab:switch_comparison_table}
	\begin{tabular}{lccr} 
		\hline
		Microwave Switch & Insertion Loss & Isolation & Return Loss\\
		\hline
		MSP6TA-12+ & 0.1 dB & 100 dB & 32.3 dB\\
		JSW6-33DR+ & 0.5 dB & 60 dB & 30.5 dB\\
		\hline
	\end{tabular}
\end{table}

\subsection{System Modelling and Calibration}
\label{sec:simulation}
It is not easy to analytically express the impact of non-ideal surface-mounted microwave switch performance and VNA measurement errors on 21-cm signal observations \citep{universe10060236}, so we developed a process to assess the effects by generating mock datasets based on actual electrical specifications of surface-mounted switches, performing calibration and sky temperature recovery, and thus calculating the foreground fitting residuals.

\subsubsection{Mock spectrum generation}
\label{sec:mock_spec_gen}
When the isolation of the microwave switch is not perfect, Equation~(\ref{eq:nw_psd}) is inadequate for its description. More generally, the receiver can be modeled as a multi-port microwave network, connecting the antenna, multiple calibrators, an LNA, and two ports of the SP6T2. When measuring a particular source, the PSD function can be expressed as
\begin{equation}  
    \begin{aligned}
    P_{\text{source}} = g_{\text{sys}}\left(\sum_{s}K_sT_s+K_{\text{amb}}T_{\text{amb}}+\sum_{\text{nw}}K_{\text{nw}}T_{\text{nw}}\right)
    \end{aligned}
    \label{eq:multiport_nw_psd}
\end{equation}
The terms in the brackets correspond to the contributions of the source temperature, the leakage of ambient temperature, and the noise wave parameters of the LNA, respectively. The receiver gain $g_{\text{sys}}$ can be determined by calibration across the entire frequency band, and no additional radiometric noise is added.
Here $T_s$ is the source temperature, for most types of calibrators it is the physical temperature, and for the antenna is the sky temperature. The coefficient of the source temperature $K_s$ is calculated using the Python package scikit-rf \citep{scikit-rf}. By modeling the source switching sub-system as a multi-port microwave network, we can calculate the power received at the LNA port originating from various sources.

$T_{\text{amb}}$ is the ambient temperature. The coefficient of the ambient temperature leakage $K_{\text{amb}}$ is calculated by the method introduced in \citet{wedge1991noise}. In this simulation, we have assumed that the switches and the two cables have the same temperature, which is a reasonable assumption given the small size of the satellite and its thermostatic regulation. 

$T_{\text{nw}}$ are the set of noise wave parameter of the LNA. The corresponding coefficients $K_{\text{nw}}$ are calculated by treating the multi-port network and all sources as an equivalent antenna, then substituting the reflection coefficient of this antenna into the coefficients of $T_{\text{unc}}$, $T_{\text{cos}}$, $T_{\text{sin}}$ in equation~(\ref{eq:nw_psd}).

\begin{figure}
	\includegraphics[width=\columnwidth]{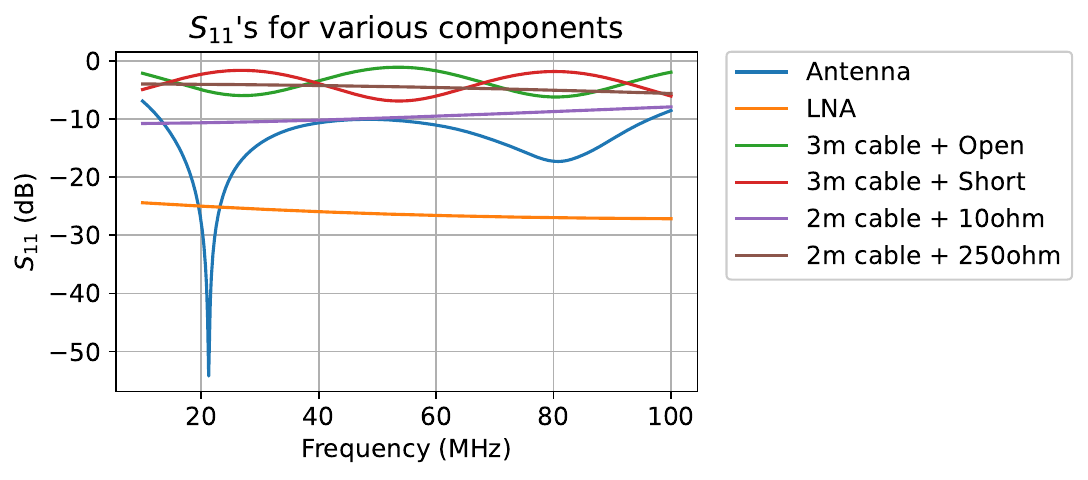}
    \caption{Reflection coefficients of various components, including antenna, LNA and calibrators, used to generate the mock data. These data are smoothed and free of noise.}
    \label{fig:simulation_S11_raw}
\end{figure}

The frequency band of the simulation is 10-100 MHz. The design of the antenna will be detailed in a subsequent paper, here we merely note that it is a foldable matched antenna. In this simulation we scale the reflection coefficient of the REACH dipole antenna \citep{de2022reach} into the observation frequency band of this instrument, as shown in Figure~\ref{fig:simulation_S11_raw}. Similarly, the same LNA as REACH is used in this simulation. The impedance of the noise source, hot load and ambient load are all assumed to be 50 $\Omega$. The reflection coefficients of other calibrators consisting of cables and terminals are obtained from advanced design system simulations, also shown in Figure~\ref{fig:simulation_S11_raw}.

We have selected several commercial microwave switches with relatively good RF performance within the observation frequency band for the simulation. The models and parameters of the microwave switches are shown in Table~\ref{tab:switch_param_table}. These parameters are flat in the observation frequency band, so they are set to be frequency-independent in the simulation.

\begin{table}
	\centering
	\caption{The RF electrical specifications of SPDT, SP6T, DPDT microwave switches used in the simulation. All data are based on typical results from measurements taken within 10-100 MHz, at 25$^{\circ}$C ambient temperature and under normal supply voltage, sourced from the official data sheet.}
	\label{tab:switch_param_table}
	\begin{tabular}{lllll} 
		\hline
		Type & Model & Insertion Loss & Isolation & Return Loss\\
		\hline
            SPDT & HMC784A    & 0.26 dB & 67 dB & 34.0 dB\\
		    SP6T & JSW6-33DR+ & 0.5 dB & 60 dB & 30.5 dB\\
            DPDT & CMD272P3   & 1.0 dB & 52 dB & 20.5 dB\\
		\hline
	\end{tabular}
\end{table}

\begin{figure}
	\includegraphics[width=\columnwidth]{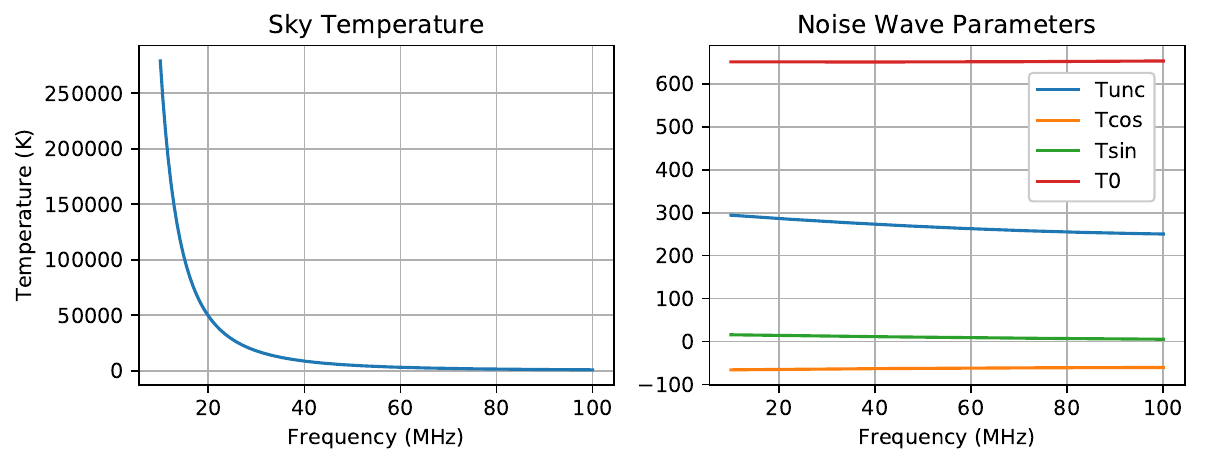}
    \caption{Left: Sky temperature considering only the Galactic synchrotron foreground. Right: The noise wave parameters of the LNA used in mock data generation.}
    \label{fig:simulation_LNA_nw}
\end{figure}

The physical temperature of the hot load is 370 K, while all other components are at 300 K. The noise source has a flat excess noise temperature of 1100K across the band. The sky temperature is dominated by the Galactic synchrotron radiation, which is a smooth power-law-like spectra. We utilize a five-term polynomial used in \citet{bowman2018absorption} to model the sky foreground temperature:
\begin{equation}  
    \begin{aligned}
    T_{\mathrm{F}}(\nu)=\sum_{n=0}^{4} a_n \nu^{n-2.5}
    \end{aligned}
    \label{eq:Sky_foreground}
\end{equation}
Here $a_n$ are the coefficients fitted to the measurement data from EDGES \citep{bowman2018absorption}, which will be used to extrapolate the foreground temperature to 10-100 MHz as the antenna temperature in the simulation, shown in Fig~\ref{fig:simulation_LNA_nw}. In this simulation, we do not consider the 21-cm signal because it is weak compared to the Galactic synchrotron signal, and our main objective is to study the effect of surface-mounted switches and measurement error on the antenna temperature recovery. The noise wave parameters of LNA for the mock data generation are also plotted in Fig~\ref{fig:simulation_LNA_nw}, which are derived from the REACH LNA noise parameter measurements in the laboratory and fitted with second-order polynomials. 
If there are more complex structures in the actual noise wave parameters once the whole receiver being assembled, we will observe it in the analysis, and then use higher-order model for the system calibration.

\subsubsection{S-parameter and temperature correction}
\label{sec:correction}
The OSL calibration are performed with respect to a reference plane of the VNA at the ports of SP6T2, while the signal paths used to measure $S_{\text{11}}$ of the antenna and the LNA include additional paths through the DPDT switch, as shown by the solid red lines  in Figure~\ref{fig:receiver_block_diagram}. However, when measuring the spectrum, the signal passes through the DPDT switch along the solid blue line. Therefore, the measured $S_{\text{11}}$ needs to be corrected for these differences. In Figure~\ref{fig:receiver_block_diagram}, the reference plane for defining the reflection coefficients of the antenna and the receiver is labelled at the end of the solid blue line near the LNA. To make this translation, we model the difference in signal paths as two-port networks, characterized by the scatter matrix [S], which can be measured in laboratory with high accuracy using the Through-Reflection-Line method\citep{rytting2001network}. The mock measurement $S_{\text{11}}$ data are generated by calculating the ratio of the output and incident voltages at the ports of the SP6T2 switch using the multi-port network model described in Section~\ref{sec:mock_spec_gen}, then corrected as \citep{pozar2011microwave},
\begin{equation}  
    \begin{aligned}
    \Gamma_{\text{out}}=S_{\text{22}}+\frac{S_{\text{12}} S_{\text{21}} \Gamma_{\mathrm{S}}}{1-S_{\text{11}} \Gamma_{\mathrm{S}}},
    \end{aligned}
    \label{eq:gamma_out}
\end{equation}
where $\Gamma_{\mathrm{S}}$ and $\Gamma_{\mathrm{out}}$ are respectively the reflection coefficients before and after embedding the DPDT .

\begin{figure}	\includegraphics[width=\columnwidth]{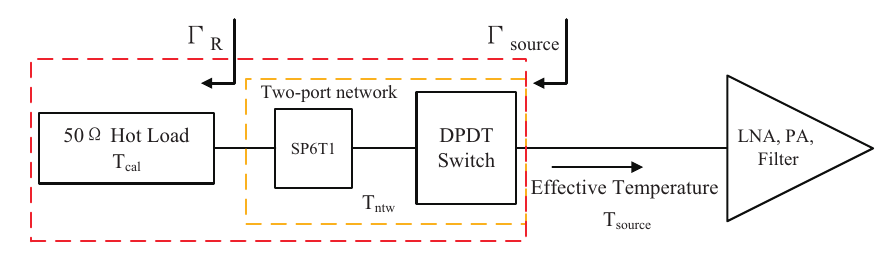}
    \caption{A schematic diagram of temperature correction for 50 $\Omega$ hot load. all switches and cables between the 50 $\Omega$ load and the LNA are treated as a two-port network. In the simulation, it is assumed that the temperature of each component within this two-port network is the same, denoted as $T_{\mathrm{ntw}}$, which is different from the temperature of the hot load, $T_{\mathrm{cal}}$.}
    \label{fig:temp_correction}
\end{figure}

During calibration, the signal path from the calibrators to the S-parameter reference plane can be treated as a two-port network for temperature correction. Figure~\ref{fig:temp_correction} illustrates this model for calculating the effective temperature of the hot load, applicable to all other calibrators and antenna. The available power gain for this two-port network is the ratio of the power available from the network to the power available from the source, expressed using the following equation in \citet{pozar2011microwave}:
\begin{equation}  
    \begin{aligned}
    G=\frac{\left|S_{21}\right|^2\left(1-\left|\Gamma_{\mathrm{R}}\right|^2\right)}{\left|1-S_{11} \Gamma_{\mathrm{R}}\right|^2\left(1-\left|\Gamma_{\text {source }}\right|^2\right)}
    \end{aligned}
    \label{eq:G_avail}
\end{equation}
where $S_{\mathrm{21}}$ and $S_{\mathrm{11}}$ are the forward S-parameters of this two-port network. $\Gamma_{\mathrm{R}}$ is the reflection coefficient of the calibrator and $\Gamma_{\mathrm{source}}$ is the reflection coefficient measured at the reference plane. Due to the non-negligible insertion loss, the  microwave switches also contribute to the noise power. With available gain, we can get the effective temperature $T_{\text {source}}$ of each source as:
\begin{equation}  
    \begin{aligned}
    T_{\text {source}}=G T_{\mathrm{cal}}+(1-G) T_{\mathrm{ntw}}
    \end{aligned}
    \label{eq:T_equivalent}
\end{equation}
where $T_{\mathrm{cal}}$ is the physical temperature of calibrator and $T_{\mathrm{ntw}}$ is the physical temperature of the two-port network. Most components would have nearly the same temperature, there are exceptions, for example the 50 $\Omega$ hot load and the antenna. In Section~\ref{sec:mock_spec_gen}, we assume a common temperatures for the switches and cables, and use a multi-port network to calculate the ambient temperature leakage.

\subsubsection{Calibration steps}
In simulation, we used different configurations, i.e. different model of switches performance and measurement errors. For each configuration, we generated a corresponding mock dataset. Each mock dataset contains:
\begin{enumerate}
 \item The spectrum of each source as measured by the receiver.
 \item Reflection coefficients of the antenna, calibrators and LNA measured at the reference plane.
 \item The S-parameters of the two-port networks representing different signal propagation paths within DPDT switch, which are used for the S-Parameter correction.
 \item The S-parameters of the two-port networks representing signal propagation paths from different sources, which are used for the temperature correction.
 \item The physical temperatures of the calibrators, cables and microwave switches.
\end{enumerate}

The calibration process is the same for each simulated dataset, listed below:

\begin{enumerate}
 \item Correct the S-parameters of different sources and LNA using the method described in Section~\ref{sec:correction}.
 \item Build the temperature correction model for different sources using the method described in Section~\ref{sec:correction}. Correct the hot load temperature, whilst the switches and cables are assumed to have a common ambient temperature.
 \item Calculate the X-terms of equation~(\ref{eq:nw_psd_simple}) and perform polynomial fit with the least squares method to calculate the noise wave parameters of the LNA. 
 \item Substitute the noise wave parameters into the antenna's PSD equation to get the antenna temperature, which is then fed into the antenna's temperature correction model to recover sky temperature. 
 \item Using Equation~(\ref{eq:Sky_foreground}) to fit the recovered sky temperature and obtain the residual, which are the systematic errors which contaminate the 21-cm signal.
\end{enumerate}

\subsection{Simulation Results}
\label{sec:results}
\subsubsection{Simulation for the effect of isolation}

In order to explore the effects of isolation on the calibration results, we generated four different datasets whose differences are listed in the Table~\ref{tab:mock_dataset_information}. In \textit{RefSet}, all microwave switches are ideal-- there is no insertion loss, reflection loss, or signal leakage—serving as a reference to validate the calibration approach. In \textit{IsoRefSet}, the insertion loss and reflection loss follow the parameters specified in Table~\ref{tab:switch_param_table}, while isolation remains ideal. In the two remaining cases \textit{SwitchOffSet} and \textit{SwitchOnSet}, we consider the impact of non-ideal isolation. When the SP6T1 in Figure \ref{fig:receiver_block_diagram} is switched to internal calibrators, due to the non-ideal isolation of the switch, there could still be some leakage caused by antenna through it to the LNA. In \textit{SwitchOffSet}, we consider the case where the SPDT1 switch is connected to the antenna when not in use, while in \textit{SwitchOnSet}, we consider the case where it switches to a 50 $\Omega$ load. This comparison is intended to demonstrate the impact on the system when the SPDT1 switch enhances the isolation of the antenna signal propagation path. All other aspects of these datasets are identical.

\begin{table}
	\centering
    \setlength{\tabcolsep}{1.2mm}
	\caption{Parameter differences between the four mock datasets. The first column is the name of the mock datasets. The second to fourth columns are the insertion loss, return loss and isolation of the microwave switches, respectively. The '-' sign indicates that the effect of this parameter is not considered, i.e., the switches have ideal insertion loss, reflection loss and isolation. The '+' sign indicates that the corresponding parameter in the datasheet is used for simulation. The fifth columns indicates whether SPDT1 switches to a 50 ohm load when the receiver measures other sources except the antenna. The '-' sign indicates not switching while '+' sign indicates switching.}
	\label{tab:mock_dataset_information}
	\begin{tabular}{c|c|c|c|c} 
		\hline
		Dataset &  Insertion Loss & Return Loss & Isolation & SPDT1 switch\\
		\hline
            \textit{RefSet} & - & - & - & -\\
		    \textit{IsoRefSet} & + & + & - & -\\
            \textit{SwitchOffSet} & + & + & + & -\\
            \textit{SwitchOnSet} & + & + & + & +\\
		\hline
	\end{tabular}
\end{table}

\begin{figure}
	\includegraphics[width=\columnwidth]{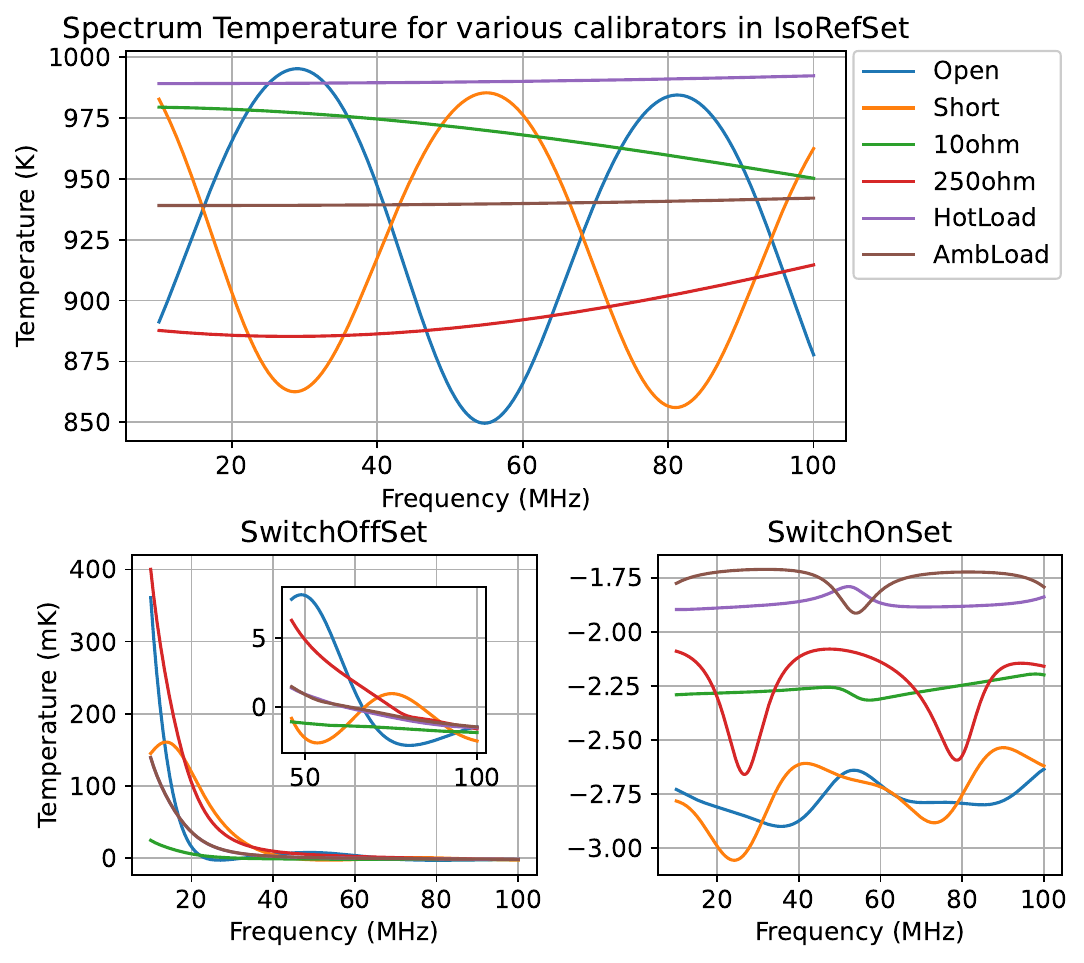}
    \caption{Top: Spectrum temperature (kelvin) for various calibrators in \textit{IsoRefSet}. Bottom left: The difference between the spectrum temperature (millikelvin) in \textit{SwitchOffSet} and that of \textit{IsoRefSet}. Bottom right: The difference between the spectrum temperature (millikelvin) in \textit{SwitchOnSet} and that of \textit{IsoRefSet}.}
    \label{fig:simulation_spec}
\end{figure}

First, we compare the spectrum of these datasets to show the impact of signal leakage. The top plot in Figure~\ref{fig:simulation_spec} shows the spectrum of various calibrators in \textit{IsoRefSet}, containing contributions from both the source temperature and LNA noise temperature.
The bottom panel illustrates the difference between the spectrum in \textit{SwitchOffSet} and \textit{SwitchOnSet} compared to \textit{IsoRefSet}, showing the effect of non-ideal isolation. If SPDT1 is kept connected to the antenna, signal leakage contributes up to 400 mK to the spectrum. If SPDT1 is switched to the $50 \Omega$ load, the signal leakage is reduced to about 3 mK, indicating that SPDT1 effectively enhance the isolation of the antenna signal propagation path.

\begin{figure}
	\includegraphics[width=\columnwidth]{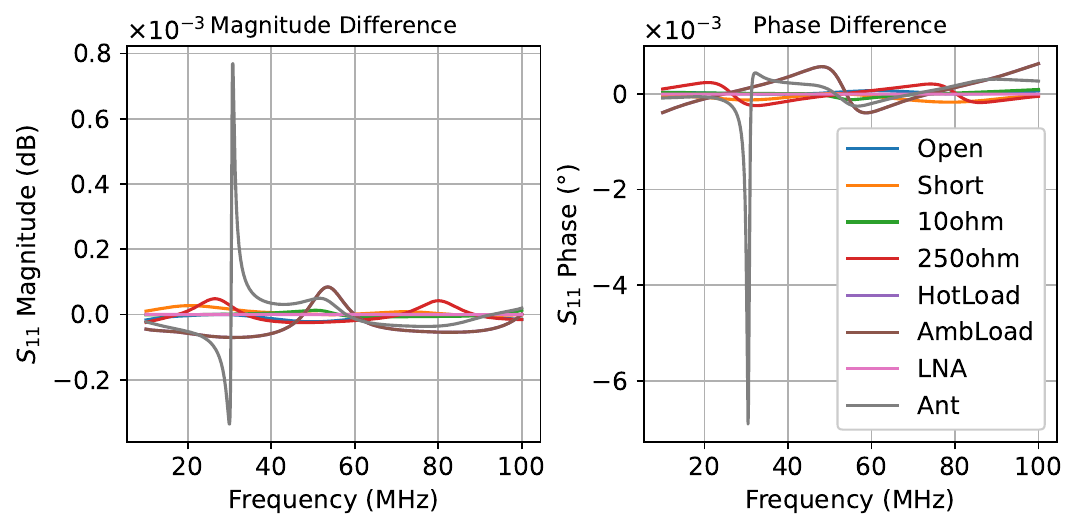}
    \caption{The difference between the corrected $S_{\text{11}}$ in \textit{SwitchOnSet} and that of \textit{IsoRefSet}. Left graph shows the magnitude difference of $S_{\text{11}}$ in dB. Right graph shows the phase difference of $S_{\text{11}}$ in degrees.}
    \label{fig:simulation_S_param_diff}
\end{figure}

Similarly, we study the effect of non-ideal isolation on the measurement of reflection coefficients for various components. Figure~\ref{fig:simulation_S_param_diff} shows the difference between the corrected $S_{\text{11}}$ in \textit{SwitchOnSet} and \textit{IsoRefSet}. The greatest impact of isolation on the measurement of S-parameters occurs at the resonance frequency of the antenna, with a magnitude of $8\times10^{-4}$ dB and a phase of $6\times10^{-3}$ degrees. It is negligible compared to the potential VNA measurement inaccuracies. Therefore, the effect of isolation on the S-parameters measurement is not significant.

\begin{figure}
	\includegraphics[width=\columnwidth]{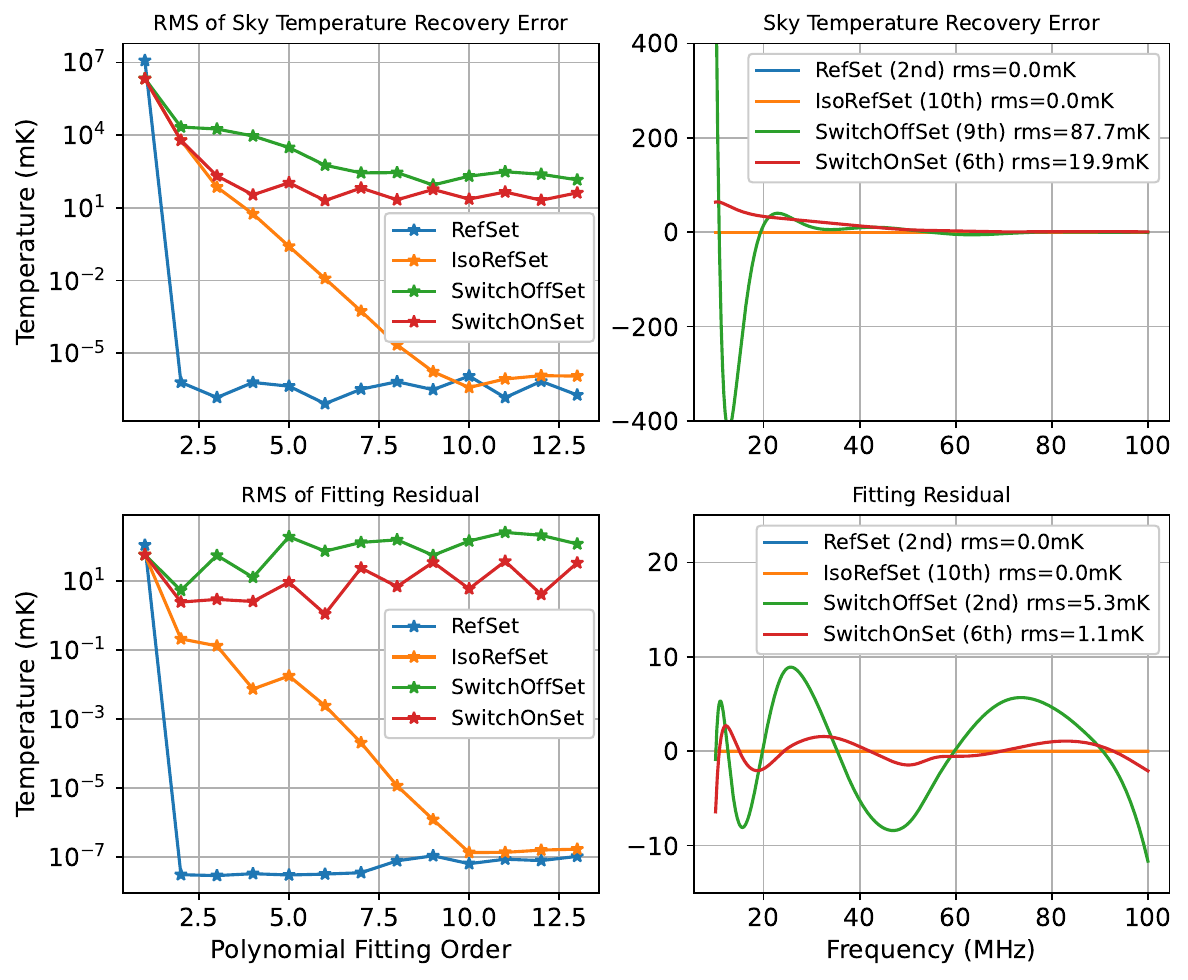}
    \caption{Left: The relationship between the RMS of the sky temperature recovery error (top) or fitting residuals (bottom) and the polynomial orders used to fit noise wave parameters in different mock datasets. Right: The recovered sky temperature error (top) or fitting residuals (bottom) for all datasets using the optimal polynomial order, respectively.}

    \label{fig:simulation_cal_result}
\end{figure}

We used polynomial functions of different orders to fit the noise wave parameters, then recover the sky temperature and get the foreground fitting residuals. Basically, we want both the recovery error and the fitting residuals to be as small as possible. Figure~\ref{fig:simulation_cal_result} shows the root mean square (RMS) and frequency structure of these two indicators. As expected, in \textit{RefSet}, the sky temperature can be perfectly recovered by using second-order polynomial functions to fit the noise wave parameters, the same order as used to generate the mock data. For \textit{IsoRefSet}, a higher-order polynomial is needed to fully recover the sky temperature. The reflection coefficient of the noise source measured at the reference plane is no longer zero when considering non-ideal switches, so the equivalent temperature of the noise source will vary with frequency. The graph shows that a $5^{\text{th}}$-order polynomial fitting results in recovery error less than 1 mK, which is sufficiently small  compared to the 21-cm signal. However, in the \textit{SwitchOffSet} and \textit{SwitchOnSet} cases, the recovery error cannot be reduced to a very low level. For \textit{SwitchOffSet}, the recovery error obtained by fitting the noise wave parameters with $10^{\text{th}}$-order polynomial is above 400 mK at low frequencies, where the sky temperature is particularly high. For \textit{SwitchOnSet}, the maximum value of the recovery error obtained by 4th-order polynomial fit is less than 60 mK. 

In actual experiments, we do not know the true sky temperature, so we are more concerned with foreground fitting residuals as it may contaminate the 21-cm signal. For \textit{SwitchOffSet}, the fitting residuals are approximately within $\pm 10$ mK at the optimal polynomial fitting order, while for \textit{SwitchOnSet}, they are within $\pm 5$ mK. Higher-order polynomials lead to higher fitting residuals because they absorb the undesirable leakage of non-ideal isolation of the switches into the noise wave parameters, improving calibrator fits but increasing sky temperature recovery errors. Considering both indicators, \textit{SwitchOffSet} can achieve low fitting residuals with $2^{\text{nd}}$-order polynomial but the recovered sky temperature significantly differs from the actual temperature. Overall, the presence of the SPDT1 switch aids in achieving smaller recovery error and fitting residuals.

\subsubsection{Simulation for the effect of measurement error}

The previous simulation assumes that all measurements are accurate, however the main sources of actual calibration error are the various measurement errors. From equation~(\ref{eq:cal_equation}), calibrating the receiver requires the measurement of three variables: PSDs, the physical temperatures and reflection coefficients, each with potential errors. We add these measurement errors to \textit{SwitchOnSet} to evaluate if the resulting foreground fitting residuals meet our requirements.

\begin{figure}
	\includegraphics[width=\columnwidth]{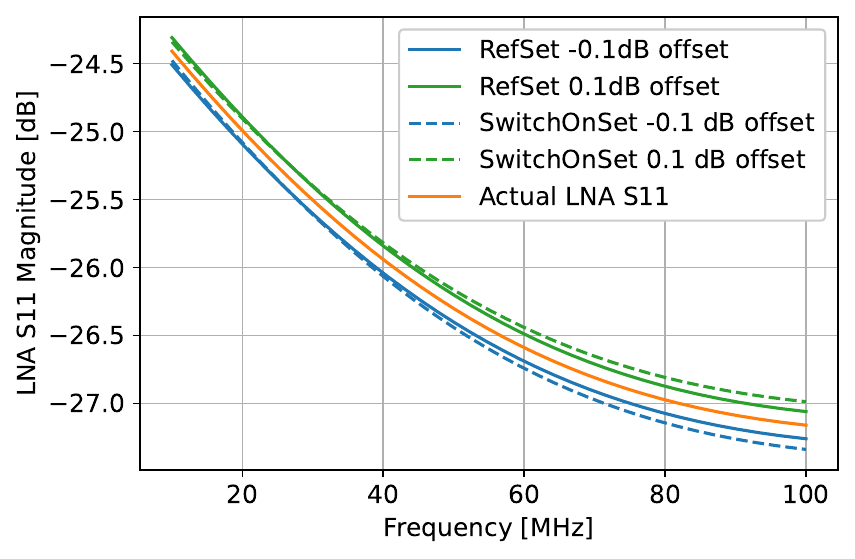}
    \caption{The corrected $S_{\text{11}}$ magnitude of the LNA when there is a measurement error of $\pm0.1$ dB at the calibration plane of the VNA. The solid line represents the results in \textit{RefSet}, while the dashed line represents the results in \textit{SwitchOnSet}.}
    \label{fig:simulation_LNAS11_error}
\end{figure}

The PSDs are measured by RFSoC-based digital spectrometer. As demonstrated in section~\ref{sec:digital}, the dynamic range is sufficient to ensure the linearity of the received PSD. Therefore, PSD measurement errors are not considered in further analyses. For physical temperature measurement, existing commercial probes can achieve an accuracy of less than $\pm$0.1 K. Based on Section~\ref{sec:vna_test} tests, we assume for calibrators connected two cables, the $S_{11}$ errors are $\pm$0.01 dB and $\pm0.2^{\circ}$. For LNA, hot load and ambient load, the $S_{11}$ error is $\pm$0.1 dB and $\pm0.2^{\circ}$. For other calibrators, the $S_{11}$ error is $\pm$0.01 dB and $\pm0.1^{\circ}$. The measurement error occurs at the calibration plane of the VNA, so they will be amplified by the imperfect S-parameters of the DPDT switch in the reference plane. Figure~\ref{fig:simulation_LNAS11_error} shows that a $\pm0.1$ dB error in the LNA's reflection coefficient measured at the VNA calibration plane can introduce additional errors of up to 0.07 dB at 100 MHz after shifting the calibration plane.

\begin{figure*}
	\includegraphics[width=2\columnwidth]{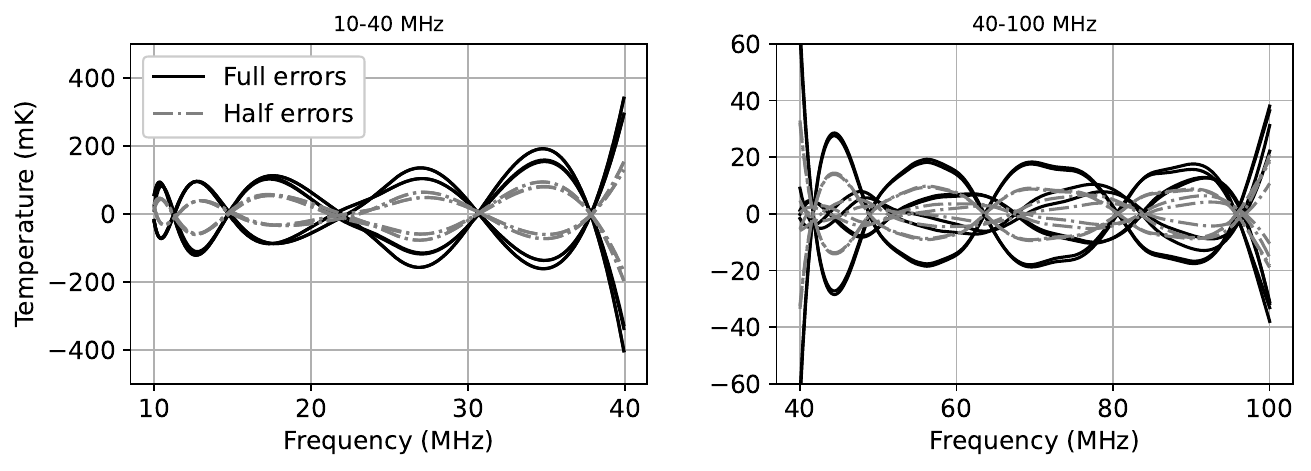}
    \caption{The fitting residuals at the optimal polynomial fitting order in \textit{SwitchOnSet} when considering all types of measurement errors. The frequency band is divided into 10-40 MHz (left) and 40-100 MHz (right). Black lines represent results with full errors, while grey lines represent results with half errors.}
    \label{fig:simulation_LNAoffset_cal_result}
\end{figure*}

Since the signs of errors in actual measurements are uncertain, we assign different error signs to the $S_{11}$ magnitude, $S_{11}$ phase and physical temperature measurements, then observe the foreground fitting residuals in different cases. As before, we calculate the RMS for different polynomial fitting orders and select the order with the lowest RMS to show the residual frequency structure. Given the large 10-100 MHz bandwidth, we divide it into two sub-bands to observe residual variations. The results are the black lines in the Figure~\ref{fig:simulation_LNAoffset_cal_result}. In 10-40 MHz range, the foreground fitting residuals are within $\pm300$ mK except for the edge of the band. It is relatively large due to high foreground temperature in low frequency, making foreground fitting much more difficult, which needs to be improved in the future. In 40-100 MHz range, which is of more interest for ground-based experiments, the recovery errors are within $\pm40$ mK except at the band edges, and up to $\pm20$ mK in 50-90MHz. The frequency structure of the residuals are related to the foreground model we are using. We also try to introduce half of the measurement error, shown as grey lines, resulting in fitting residuals being halved as well.

\begin{figure*}
	\includegraphics[width=2\columnwidth]{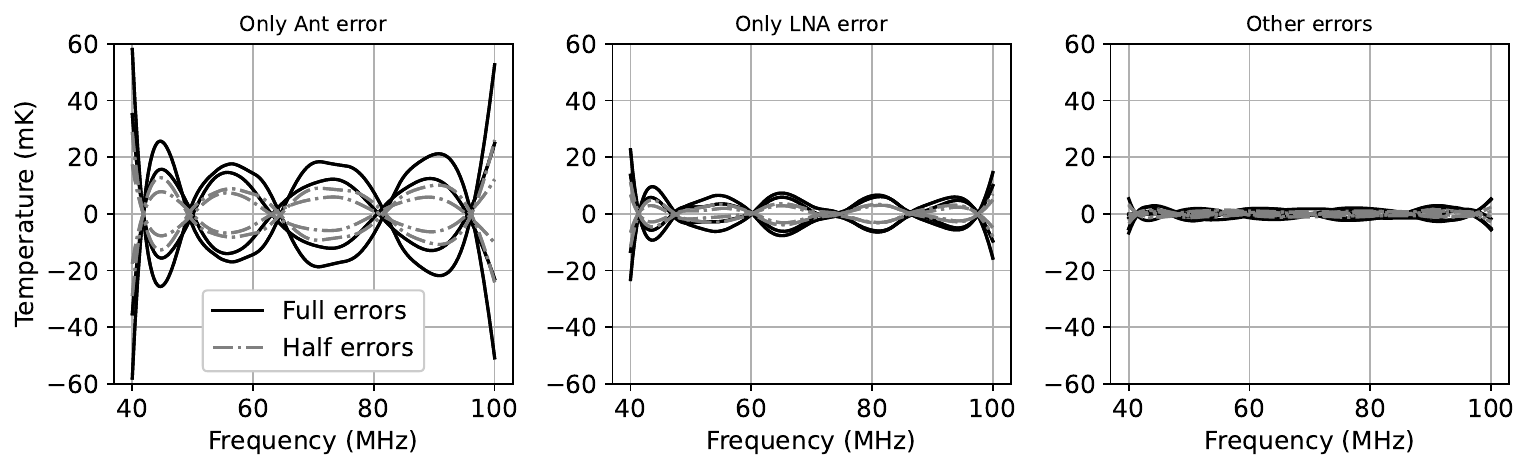}
    \caption{The fitting residuals at the optimal polynomial fitting order in \textit{SwitchOnSet} when considering different types of measurement errors in 40-100 MHz. The Left plot considers only the antenna's $S_{11}$ measurement error, the middle plot considers only LNA's $S_{11}$ measurement error, and the right plot accounts for all other $S_{11}$s and physical tempearture measurement errors. Black lines represent results with full errors, while grey lines represent results with half errors.}
    \label{fig:simulation_LNAoffset_sep_cal_result}
\end{figure*}

Figure~\ref{fig:simulation_LNAoffset_sep_cal_result} illustrates the contribution of different measurement errors to the final fitting residuals in the 40-100 MHz range. The antenna's $S_{11}$ measurement error is the primary source of residual, followed by the LNA's $S_{11}$ error, while other errors have relatively small contribution to the final fitting residuals.

\section{Conclusions and future work}
\label{sec:conclusion}
CosmoCube plans to perform global 21-cm spectrum experiment in lunar orbit to avoid RFI, ground and ionosphere effects. The receiver design is based on RFSoC, which integrates RF data converters, FPGA, CPU and several peripheral interfaces on a single board, helping to reduce size, weight and power consumption. We presented the measurement principles and design of the RFSoC-based VNA module, discussed the effect of quantization errors on reflection coefficient measurements using the MATLAB Simulink Platform. The main error arises from low-power measurements on DUTs with low reflection coefficients. We performed laboratory tests on the VNA module, including minimum output power, accuracy, stability and trace noise. The largest measurement error occurred at the resonance frequency of the filter where the reflection coefficient changed dramatically. In addition, when calibrating and measuring a 13 dB attenuator terminated with an Open calibration standard at -25 dBm output power level, the measurements differed from the PNA results in the laboratory by approximately 0.1 dB and $0.1^{\circ}$. We also presented the design of the source switching sub-system. Since it is necessary to use surface-mounted switches, we generated different mock datasets and discussed the effect of the RF electrical performance of these switches on the sky temperature recovery and foreground fitting residuals through simulation. The effect of signal leakage can be reduced by adding a microwave switch to improve the isolation of the antenna signal path. Foreground signals can be accurately subtracted if all physical temperatures and reflection coefficients are accurately measured. Considering all types of measurement errors, the fitting residuals of the foreground in 50-90 MHz can be controlled within $\pm20$ mK, small enough compared to 21-cm signal. The antenna' $S_{11}$ measurement error is the primary source of foreground fitting residual.

Future work will focus on how to improve the measurement accuracy of the VNA module, especially at the resonance point. For the source switching sub-system, it is still necessary to search better methods to ensure signal isolation, as well as to find more suitable cable lengths and impedances of calibrators to improve system calibration accuracy. Since the VNA measurement error is inevitable, We will try to find a better method that takes this error into account to improve the calibration accuracy and subtract the sky foregroundas much as possible, especially at low frequencies.

\section*{ Acknowledgements}
Jiacong Zhu acknowledges the Chinese Scholarship Council (CSC) for the funding. Eloy de Lera Acedo acknowledges STFC for their funding support through a STFC Ernest Rutherford Fellowship. Xuelei Chen acknowledges the support of the Chinese Academy of Science grant ZDKYYQ20200008 and the National Natural Science Foundation Grant 12361141814.

\section*{ Data Availability}

The data underlying this article will be shared on reasonable request to the corresponding author.



\bibliographystyle{rasti}
\bibliography{refs}

\appendix

\section{Abbreviations}
Some important abbreviations used in the text and their definitions are listed in Table~\ref{tab:abbreviations}.
\begin{table}
\centering
\caption{Abbreviations used in the text and their definitions.}
\label{tab:abbreviations}
\begin{tabular}{ll} 
    \hline
    Abbreviations & Definition\\
    \hline
        RFSoC &  Radio Frequency system-on-chip \\
        VNA  & Vector Network Analyzer \\
        RFI & Radio Frequency Interference \\
        LNA  & Low Noise Amplifier  \\
        ADC & Analog-to-digital Converter \\
        DAC & Digital-to-analog Converter \\
        GSPS & Gigabit Samples Per Second \\
        FPGA & Field Programmable Gate Array \\
        SMA & SubMiniature Version A \\
        SFDR & Spurious Free Dynamic Range \\
        PSD & Power Spectral Density \\
        NCO & Numerically Controlled Oscillator \\
        DC & Direct Current \\
        DUT & Device Under Test \\
        OSL & Open-Short-Load \\
        RMS & Root Mean Squares\\
        DPDT & Double Pole Double Throw \\
        SPDT & Single Pole Double Throw \\
        SP6T & Single Pole Six Throw \\
    \hline
\end{tabular}
\end{table}





\bsp	
\label{lastpage}
\end{document}